\documentclass[a4paper,11pt]{article}
\pdfoutput=1
\usepackage[utf8]{inputenc}
\usepackage{jheppub}

\usepackage[T1]{fontenc}

\usepackage{listings}

\preprint{DESY 16-093  \\ \phantom{xxx} \hfill  KCL-PH-TH/2016-30}

\title{Constraints on \texorpdfstring{$Z'$}{Z'} models from LHC dijet searches and implications for dark matter}

\author[a]{Malcolm Fairbairn,}
\author[a]{John Heal,}
\author[b]{Felix Kahlhoefer}
\author[a]{and Patrick Tunney}

\affiliation[a]{Physics, King's College London, Strand, London, WC2R 2LS, UK}
\affiliation[b]{DESY, Notkestra\ss e 85, D-22607 Hamburg, Germany}

\emailAdd{malcolm.fairbairn@kcl.ac.uk}
\emailAdd{john.heal@kcl.ac.uk}
\emailAdd{felix.kahlhoefer@desy.de}
\emailAdd{patrick.tunney@kcl.ac.uk}

\abstract{We analyse a combination of ATLAS and CMS searches for dijet resonances at run I and run II, presenting the results in a way that can be easily applied to a generic $Z'$ model.  As an illustrative example, we consider a simple model of a $Z'$ coupling to quarks and dark matter. We first study a benchmark case with fixed couplings and then focus on the assumption that the $Z'$ is responsible for setting the dark matter relic abundance. Dijet constraints place significant bounds on this scenario, allowing us to narrow down the allowed range of dark matter masses for given $Z'$ mass and width.}


\begin{document}
\maketitle
\flushbottom

\section{Introduction}
Resonant structures in the invariant mass distribution of dijet events are amongst the most generic signatures for ``exotic'' new physics at the LHC, since any new heavy particle produced in the $s$-channel at hadron colliders can decay back into a pair of jets. Searches for dijet resonances are therefore a high priority at both ATLAS and CMS and have been among the first searches carried out at a centre-of-mass energy of 13 TeV \cite{ATLAS13,CMS13}. These searches are complemented by earlier searches at 8 TeV \cite{ATLAS8,CMS8}, as well as a dedicated search for dijet resonances with an invariant mass below 1 TeV at CMS based on a novel data scouting technique~\cite{Khachatryan:2016ecr}. Among the many models probed by such searches are Randall-Sundrum (RS) gravitons~\cite{Randall:1999ee}, excited quarks~\cite{Baur:1987ga,Baur:1989kv} and models with a leptophobic $Z'$~\cite{FileviezPerez:2010gw,Dobrescu:2013coa,Chiang:2015ika} (see~\cite{Ekstedt:2016wyi} for the implications of these searches on more generic $Z'$ models).

The latter model of a massive spin-1 boson that couples predominantly to quarks has received a significant amount of interest in the context of dark matter (DM) production at hadron colliders. The reason is that a leptophobic $Z'$ can have large couplings to the DM particle and thereby mediate the interactions that keep DM in thermal equilibrium in the early Universe. We can then hope to experimentally probe these interactions with a range of different DM search experiments. Two sensitive probes of such scenarios are direct detection experiments searching for evidence of DM-nucleus scattering and searches for missing energy at the LHC~\cite{Frandsen:2012rk,Fox:2012ru,Alves:2013tqa,Arcadi:2013qia,Buchmueller:2013dya,Buchmueller:2014yoa,Lebedev:2014bba,Harris:2014hga,Busoni:2014gta,Fairbairn:2014aqa,Jacques:2015zha,Alves:2015pea,Chala:2015ama,DeSimone:2016fbz,Brennan:2016xjh,Jacques:2016dqz}. In fact, DM models with a leptophobic $Z'$ have served to inspire a class of so-called simplified DM models, which are now commonly used to optimise LHC searches for DM~\cite{Abdallah:2015ter,Abercrombie:2015wmb}.

However, as emphasised in a number of previous works~\cite{An:2012va, An:2012ue,Frandsen:2012rk,Chala:2015ama}, DM models with a massive spin-1 mediator cannot only be probed by conventional DM searches, but also by direct searches for the mediator, in particular searches for dijet resonances. Due to the presence of invisible decays of the mediator, the width of resonance may be broadened, making it harder to distinguish the signal from the smoothly-falling QCD backgrounds. The purpose of the present work is to derive combined limits on such models by re-analysing all available LHC searches for dijet resonances. 

Previous combinations of dijet constraints have either focussed on narrow resonances~\cite{Frandsen:2012rk}, resonances that decay exclusively into quarks~\cite{Dobrescu:2013coa} or on individual DM models with specific choices of couplings~\cite{An:2012va,Chala:2015ama}. The present work aims to take a largely model-independent approach, so that our results can be applied to a range of different $Z'$ models. For this purpose, we take the width of the $Z'$ as a free parameter, which we allow to be as large as $\Gamma / m_{Z'} = 0.3$. The resulting bound on the $Z'$-quark coupling can then be applied to models where the $Z'$ decays exclusively to quarks (for example if couplings to DM are negligible or absent, or if decays of the $Z'$ into DM available are kinematically forbidden), to models where the width of the $Z'$ is given exclusively in terms of its couplings to quarks and DM (as is the case in simplified DM models) and to models where additional unobserved decay channels may be present that further broaden the $Z'$ width. To illustrate our approach, we show how the resulting dijet constraints can be applied to simplified DM models. Moreover, we develop a new method to combine dijet constraints with information on the relic abundance of DM to determine those regions of parameter space where thermal freeze-out via a $Z'$ is incompatible with constraints from the LHC.
 
The outline of our work is as follows: We first perform a combined analysis of searches for dijet resonances from the CMS and ATLAS experiments at both 8 and 13 TeV in section~\ref{sec:dijets}. We present the results of this analysis in terms of an upper limit on the $Z'$-quark coupling as a function of the $Z'$ mass and width, which applies to a wide range of $Z'$ models irrespective of what other couplings and particles are present in the model. In section~\ref{sec:DM} we then show how these results may be applied to a specific model, namely a model with a $Z'$ mediator coupling to quarks and DM. We first consider a simplified model similar to the one presently employed by the LHC collaborations and then combine such a model with relic density constraints to place limits on the coupling and the mass of the DM particle. Additional technical details are provided in the appendices.

\section{Limits on generic \texorpdfstring{$Z'$}{Z'} models from dijet resonance searches at the LHC}
\label{sec:dijets}

In this section we describe our technique for the re-analysis of searches for dijet resonances from the LHC run I and II and present the resulting combined limits. Our basic approach can be divided into three separate steps. We first implement a fully general $Z'$ model in a Monte Carlo generator to produce dijet events at the relevant centre-of-mass energies and apply the selection cuts corresponding to the various analyses. We then compare the predicted distributions of the dijet invariant mass to the experimental data, employing the same strategy as the experimental collaborations to model the background contribution. Finally, we combine the experimental tension from all data sets in a statistically consistent way in order to determine the largest signal strength that can still be compatible with all experimental data. The details of each of these steps are discussed in the following subsections.

\subsection{Dijet event generation}

The first step in our analysis is to generate dijet events resulting from a new $Z'$ mediator with mass $500\:\text{GeV} \leq m_{Z'} \leq 4\:\text{TeV}$, which interacts with the standard model quarks $q$ via the vectorial coupling $g_q$. To generate this signal, we add the following terms to the SM Lagrangian:
\begin{align}
    \mathcal{L}_\text{kin} & = -\frac{1}{4} F'_{\mu \nu} F'^{\mu \nu} + \frac{1}{2} m_{Z'}^2 Z'_{\mu} Z'^{\mu} \label{eq:lag1z} \; ,\\
    \mathcal{L}_\text{int} & =- g_q Z'_\mu \sum_q \bar{q} \gamma^\mu q \label{eq:lag2z} \; ,
\end{align}
where $F'^{\mu\nu} = \partial^\mu Z'^\nu - \partial^\nu Z'^\mu$. Although we do not specify any other couplings of the $Z'$, such couplings may in principle be present.\footnote{In particular, we assume at this point that the $Z'$ has vectorial couplings to quarks. We will discuss below how our results can be applied to models with axial couplings to quarks and to a combination of vectorial and axial couplings.} We therefore do not calculate the total decay width $\Gamma$ of the $Z'$ in terms of its quark coupling but instead take $\Gamma$ to be a free parameter of the model.  

An important advantage of taking $\Gamma$ and $g_q$ as independent parameters is that the shape of the dijet invariant mass distribution depends only on the two parameters $m_{Z'}$ and $\Gamma$, whereas the total magnitude of the signal is proportional to $g_q^4$ irrespective of whether the $Z'$ is produced on-shell or off-shell.\footnote{Note in particular that we allow the unphysical situation that the branching ratio into quarks, given by $\Gamma^{q\bar{q}} / \Gamma$, can become larger than unity. This does not pose a problem as long as the resulting constraints are only evaluated for physical combinations of $g_q$ and $\Gamma$.} We can therefore generate events for different values of $m_{Z'}$ and $\Gamma$ and a fixed value of $g_q$ and apply the simple rescaling $\sigma \propto g_q^4$ to obtain signal predictions for the full three-dimensional parameter space.

Our simulation of dijet events is carried out using a pipeline of the publicly available software packages \texttt{FeynRules\_v1.6.11}~\cite{Alloul:2013bka}, \texttt{MadGraph\_v3.2.2}~\cite{Alwall:2014hca}, \texttt{Pythia\_v8.186}~\cite{Sjostrand:2007gs,Sjostrand:2014zea} and \texttt{FastJet\_v3.0.5}~\cite{Cacciari:2011ma}. First, we implement the model Lagrangian in \texttt{FeynRules} to calculate the Feynman rules and generate a UFO model file~\cite{Degrande:2011ua}. In \texttt{MadGraph} we then generate matrix elements for all processes involving a virtual $Z'$ and a pair of $u$, $d$, $s$, $c$ or $b$ quarks in the final state.

The output from MadGraph is interfaced with Pythia, which we use both as a Monte Carlo event generator and to simulate showering and hadronisation. For our simulations, we use the \texttt{CTEQ5L} parton distribution function~\cite{Lai:1999wy}. We neglect next-to-leading order corrections, which are expected to lead to somewhat larger cross sections~\cite{Accomando:2010fz}, so we give a conservative bound. The resulting final states are clustered with \texttt{FastJet} using the anti-$k_T$ algorithm \cite{Cacciari:2008gp}. We choose the jet parameters (cone-size $R$, maximum pseudorapidity $\eta$ and minimum transverse momentum $p_{T \text{min}}$ of the jet) to match those adopted by each experiment under consideration (see table \ref{tab:jets}).

\begin{table}[tbp]
\centering
\begin{tabular}{ccccc}
\hline
\hline
&$R$&$|\eta|$& $p_{T \text{min}}$ & Ref.\\
\hline
ATLAS 13 TeV & 0.4 & $<2.4$ & 50 GeV & \cite{ATLAS13}\\
CMS 13 TeV & 1.1 & $<2.5$ & 30 GeV  & \cite{CMS13}\\
ATLAS 8 TeV & 0.6 & $<2.8$ & 50 GeV & \cite{ATLAS8}\\
CMS 8 TeV & 1.1 & $<2.5$ & 30 GeV & \cite{CMS8}\\
CMS 8 TeV (low $m_{jj}$) & 1.1 & $< 2.5$ & 30 GeV & \cite{Khachatryan:2016ecr} \\
\hline
\hline
\end{tabular}
\caption{\label{tab:jets} Jet parameters chosen for the anti-$k_T$ algorithm for the five experimental searches. The radius parameter is defined as $R = \sqrt{(\Delta \eta)^2 + (\Delta \phi)^2}$ with $\phi$ the azimuthal angle. For the CMS analyses, jets are first reconstructed with a radius parameter of 0.5 (0.4) at 8 TeV (13 TeV) and are then combined into two fat jets with radius parameter 1.1.} 
\end{table}

Once the jets have been reconstructed, we apply the experimental selection cuts outlined in table~\ref{tab:cuts}. For each event that passes these cuts we calculate the invariant mass of the dijet system. Rather than performing a full detector simulation, we approximate the uncertainties in reconstructing the energy and momentum of the jet events arising from detector performance by applying a Gaussian smearing to the invariant dijet mass. For this purpose, we take the detector resolution in both ATLAS and CMS to be 
\begin{equation}
\sigma(m_{jj})=1.8 {\rm GeV} \sqrt{m_{jj}/{\rm GeV}} \; ,
\label{eq:smear}
\end{equation}
which was determined by fitting the smeared signals to shapes given by the CMS experiment~\cite{CMS8} for a RS graviton benchmark model. The smeared invariant masses are then binned according to the bin sizes given by the different experiments and the resulting histograms are converted into differential cross sections $\mathrm{d}\sigma_{Z'}/\mathrm{d}m_{jj}$.

\subsection{Compatibility of a dijet signal with LHC data}

Once the dijet invariant mass distributions have been generated, the second step is to determine the compatibility of such a signal with observations at the LHC. For this purpose, we follow the approach of the experimental collaborations and assume that the SM background can be described by a smooth function of the form:
\begin{equation}
\frac{\mathrm{d}\sigma_\text{SM}}{\mathrm{d}m_{jj}}=\frac{P_0\left(1-m_{jj}/\sqrt{s}\right)^{P_1}}{\left(m_{jj}/\sqrt{s}\right)^{P_2+P_3\log\left(m_{jj}/\sqrt{s}\right)}} \; ,
\label{fitfunc}
\end{equation}
where the parameters $P_i$ are determined by fitting the function to the data.\footnote{While the CMS analyses at 8 and 13 TeV and the ATLAS analysis at 8 TeV allow all four parameters to vary, the ATLAS analysis at 13 TeV fixes $P_3=0$.} The total dijet invariant mass distribution is then given by $\mathrm{d}\sigma/\mathrm{d}m_{jj} = \mathrm{d}\sigma_\text{SM}/\mathrm{d}m_{jj} + \mathrm{d}\sigma_{Z'}/\mathrm{d}m_{jj}$, where the first term depends on the unknown parameters $P_i$, while the second term depends on the assumed values for $m_{Z'}$, $\Gamma$ and $g_q$.

To compare the model prediction to experimental data, we calculate the usual $\chi^2$ test statistic
\begin{equation}
  \chi^2=\sum_i \left(\frac{d_i - s_i}{\sigma_i}\right)^2 \; ,
\label{chisq}
\end{equation}
where the index $i$ denotes the bin number in a given experiment, $d_i$ is the observed differential cross section with corresponding (statistical) error $\sigma_i$ and $s_i$ is the predicted signal containing both the SM contribution and the new-physics signal. We now fix the unknown parameters $P_i$ by finding the minimum of the $\chi^2$ distribution with respect to these parameters (called $\hat{\chi}^2$).

\begin{table}[tbp]
\centering
\begin{tabular}{ccccc}
\hline
\hline
&$m_{jj}$  &$|\Delta \eta_{jj}|$& additional & Ref.\\
\hline
ATLAS 13 TeV & $>1.1$ TeV & $<1.2$ & $p_{T,j_1}>440$ GeV and $p_{T,j_2}>50$ GeV &\cite{ATLAS13} \\
CMS 13 TeV & $>1.2$ TeV & $<1.3$ & $p_{T,j_1} > $ 500 GeV or $H_T > 800$ GeV &\cite{CMS13}\\
ATLAS 8 TeV & $>250$ GeV & $<1.2$ & - & \cite{ATLAS8}\\
CMS 8 TeV & $>890$ GeV & $<1.3$ & - & \cite{CMS8}\\
CMS 8 TeV (low) & $> 390$ GeV & $< 1.3$ & - & \cite{Khachatryan:2016ecr}\\
\hline
\hline
\end{tabular}
\caption{\label{tab:cuts} Experimental cuts adopted by the five experimental searches. $p_{T,j_1}$ refers to the transverse momentum of the leading jet, while $p_{T,j_2}$ refers to the subleading jet. $H_T$ is the scalar sum of all jet $p_T$ for jets with $p_T>$ 40 GeV and $|\eta| < 3$, and $\Delta \eta_{jj}$ refers to the rapidity separation of the leading and subleading jets.}
\end{table}

We now want to place an upper bound on the magnitude of the new-physics signal, beyond which the sum of signal and background are incompatible with the data. For this purpose, we employ a $\Delta \chi^2$ method. We first calculate $\hat{\chi}^2$ in the absence of a contribution from the $Z'$ mediator (called $\hat{\chi}^2_0$) and then define $\Delta \chi^2(m_{Z'},\Gamma,g_q) = \hat{\chi}^2(m_{Z'},\Gamma,g_q) - \hat{\chi}^2_0$.

For $\Delta \chi^2 < 0$, the data actually prefers a non-zero contribution from the $Z'$ mediator. Positive values of $\Delta \chi^2$, on the other hand, are disfavoured by the data. In such a case, we can calculate the $p$-value, i.e.\ the probability to observe at least as large a value of $\Delta \chi^2$ from random fluctuations in the data as
\begin{equation}
P=1-{\rm CDF}\left(1,\Delta\chi^2\right)
\label{eq:chiprob}
\end{equation}
where ${\rm CDF}\left(1,\chi^2 \right)$ is the cumulative distribution function for the $\chi^2$ distribution with one degree of freedom.\footnote{We note that the $\Delta \chi^2$ test statistic as we define it does not exactly follow a $\chi^2$-distribution. The reason is that we take $\hat{\chi}^2_0$ to be the value of $\hat{\chi}^2$ for $g_q = 0$ rather than finding the value of $g_q$ that actually minimises $\hat{\chi}^2$ (called $g_{q,0}$) in order to avoid the problem that the data may prefer a negative signal contribution. Since $\hat{\chi}^2(0) \geq \hat{\chi}^2(g_{q,0})$, our definition yields slightly smaller values for $\Delta \chi^2$ than the one obtained from minimising $\hat{\chi}^2$ with respect to $g_q$. Using a $\chi^2$-distribution to calculate the $p$-value therefore means that we slightly overestimate the $p$-value and consequently place more conservative bounds. We have verified that the error made by this approximation is small by determining the actual distribution of $\Delta \chi^2$ from a Monte Carlo simulation for specific parameter points.}

As discussed above, the new-physics signal is proportional to $g_q^4$. As we increase $g_q$, keeping $m_{Z'}$ and $\Gamma$ fixed, we will reach the point where $\Delta \chi^2$ becomes so large that $P$ becomes unacceptably small. For $P < 5\%$, we can exclude the corresponding value of $g_q$ at the 95\% confidence level. If $P > 5\%$, the value of $g_q$ cannot be excluded by the experiment under consideration, but it may still be excluded by the combination of results from several experiments. Such a combination is necessarily model-dependent in the sense that it requires an assumption on the ratio of the production cross section of the resonance at 8 TeV and 13 TeV. Since we use a $Z'$-model to generate dijet events, our combination is valid for any resonance produced dominantly from light quarks (with equal couplings to all flavours).

For a given signal hypothesis, we can follow the procedure described above to obtain a value of $\Delta \chi^2$ for each experiment under consideration. Crucially, the parameters $P_i$ are fitted independently for each experiment. Since the two CMS searches at 8 TeV are not statistically independent, we use ref~\cite{CMS8} for $m_{Z'} \geq 1\:\text{TeV}$ and ref~\cite{Khachatryan:2016ecr} for smaller $Z'$ masses. We can then simply add up the individual contributions to $\Delta \chi^2$ to obtain $\Delta \chi^2_\text{total}$. This test statistic is again expected to approximately follow a $\chi^2$-distribution with one degree of freedom, so we can calculate the combined $p$-value with eq.~(\ref{eq:chiprob}). 

\subsubsection*{Validation based on the RS-graviton model}

\begin{figure}[t]
  \centering
    \includegraphics[width=0.8\textwidth]{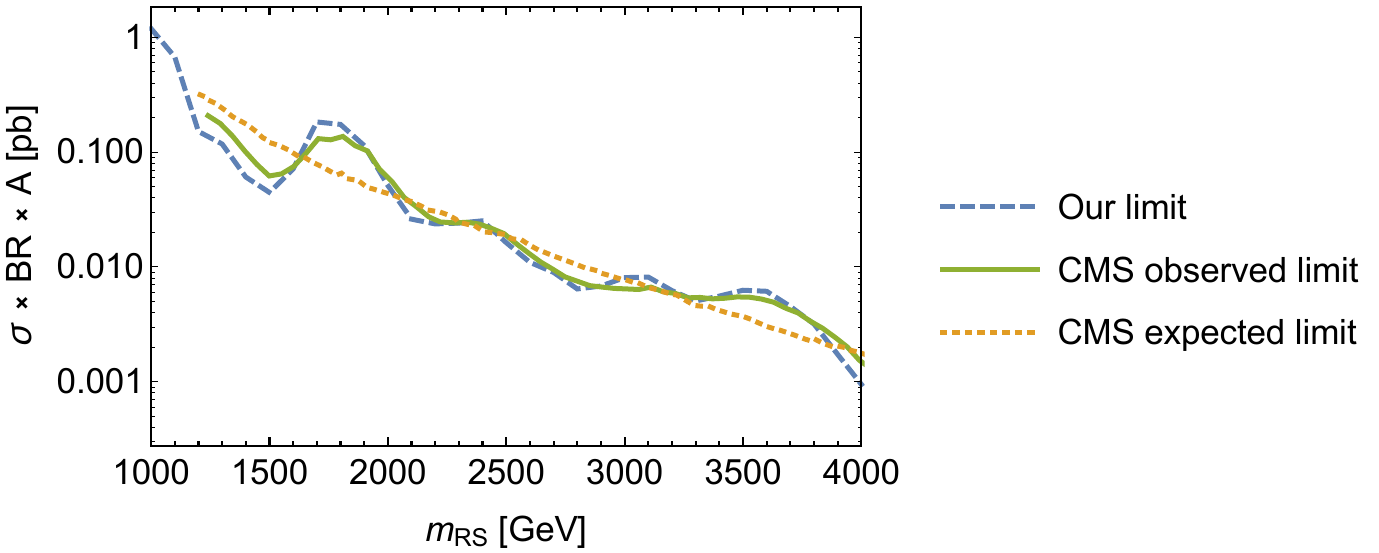}
  \caption{Comparison of our method for setting limits from dijet data (blue, dashed) with those from the CMS experiment~\cite{CMS8} (green, solid) presented in terms of the cross section times acceptance, for the benchmark model of an RS graviton~\cite{Randall:1999ee}.\label{fig:rsvalid}}
\end{figure}

To validate our limit-setting procedure, we have applied our analysis chain to the RS graviton model~\cite{Randall:1999ee}, which is used as a fiducial benchmark by CMS~\cite{CMS8}. The model contains two free parameters, namely the mass of the RS graviton $m_\text{RS}$ and the curvature of the five-dimensional bulk $k/\tilde{M_\text{Pl}}$ where $\tilde{M_\text{Pl}}$ is the reduced Planck mass. The latter is taken to be $k=0.1\tilde{M_\text{Pl}}$, which fully determines the width and the couplings relevant for the generation of dijet events for a given value of $m_\text{RS}$. We adopt these values to calculate the total production cross section and to generate dijet invariant mass distributions. We then multiply the resulting distributions with a rescaling factor $\mu$ in order to determine the largest signal strength that is still compatible with data. Applying the resulting rescaling factor to the total cross section then yields an upper bound on the production cross section as a function of the RS graviton mass. Figure~\ref{fig:rsvalid} shows the comparison of the bounds we obtain with the results from the CMS analysis. We conclude that our approach yields good agreement with the results from the CMS collaboration over a wide range of masses.

\subsection{Results}

Having described the strategy for deriving bounds on the $Z'$-quark coupling $g_q$, we now present the results of our analysis in a way that can be applied to a wide range of $Z'$ models. We consider $m_{Z'}$ masses between 500 GeV and 4 TeV in steps of 50 GeV. For each mediator mass, we consider six different widths: 1\%, 2\%, 5\%, 10\%, 20\% and 30\% (in units of $m_{Z'}$). These values were chosen to best sample the variation in the constraints for different widths. In particular, we note that for a $Z'$ width smaller than 1\% of $m_{Z'}$, the shape of the dijet invariant mass distribution is dominated by the detector resolution and therefore becomes independent of $\Gamma$.  For each combination of $m_{Z'}$ and $\Gamma$ we then determine the largest value of $g_q$ that is compatible with the experimental data at 95\% confidence level (called $g_{q,95\%}$).

\begin{figure}[t]
  \centering
 \includegraphics[width=0.45\linewidth]{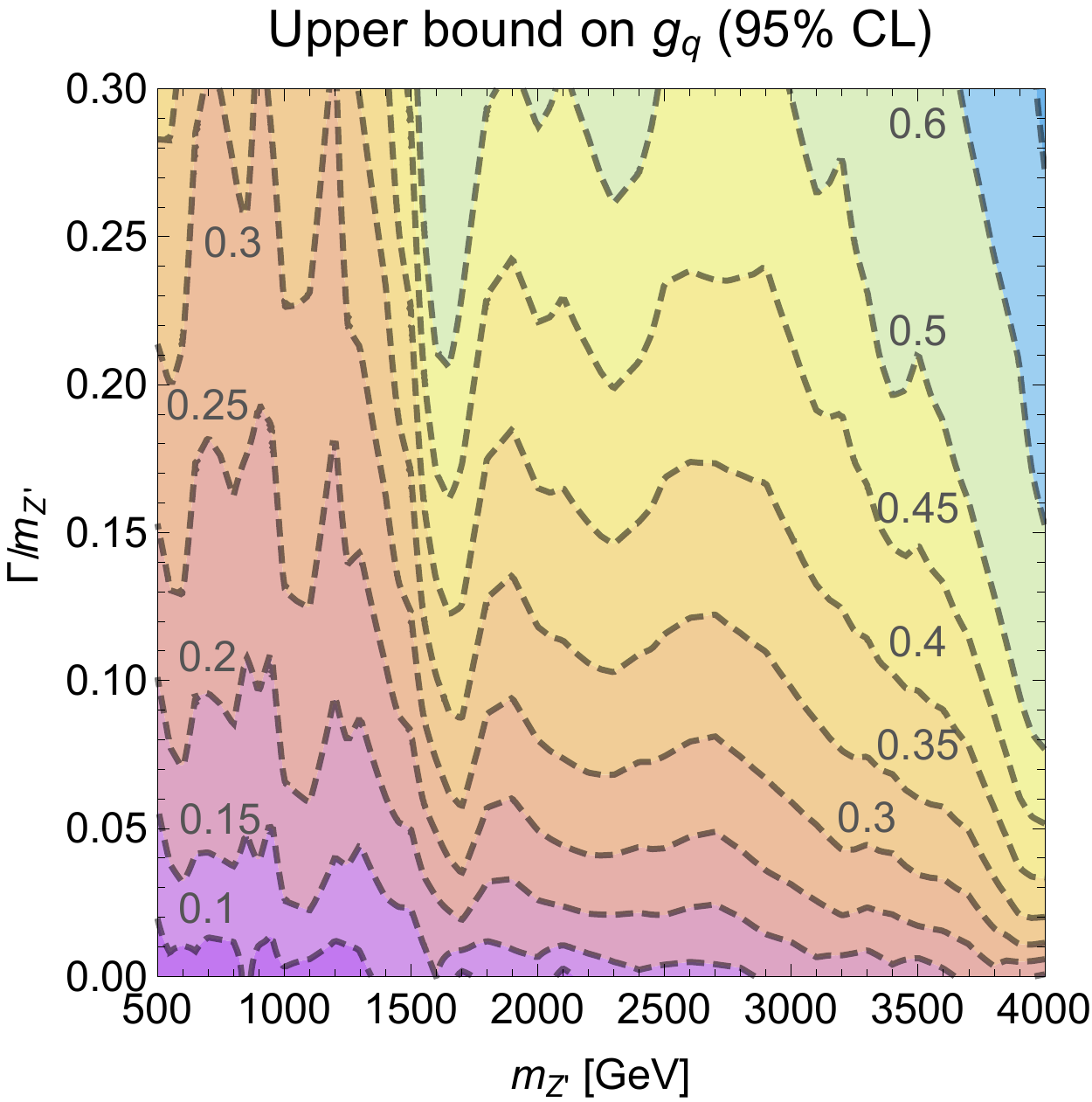}\qquad
  \includegraphics[width=0.45\linewidth]{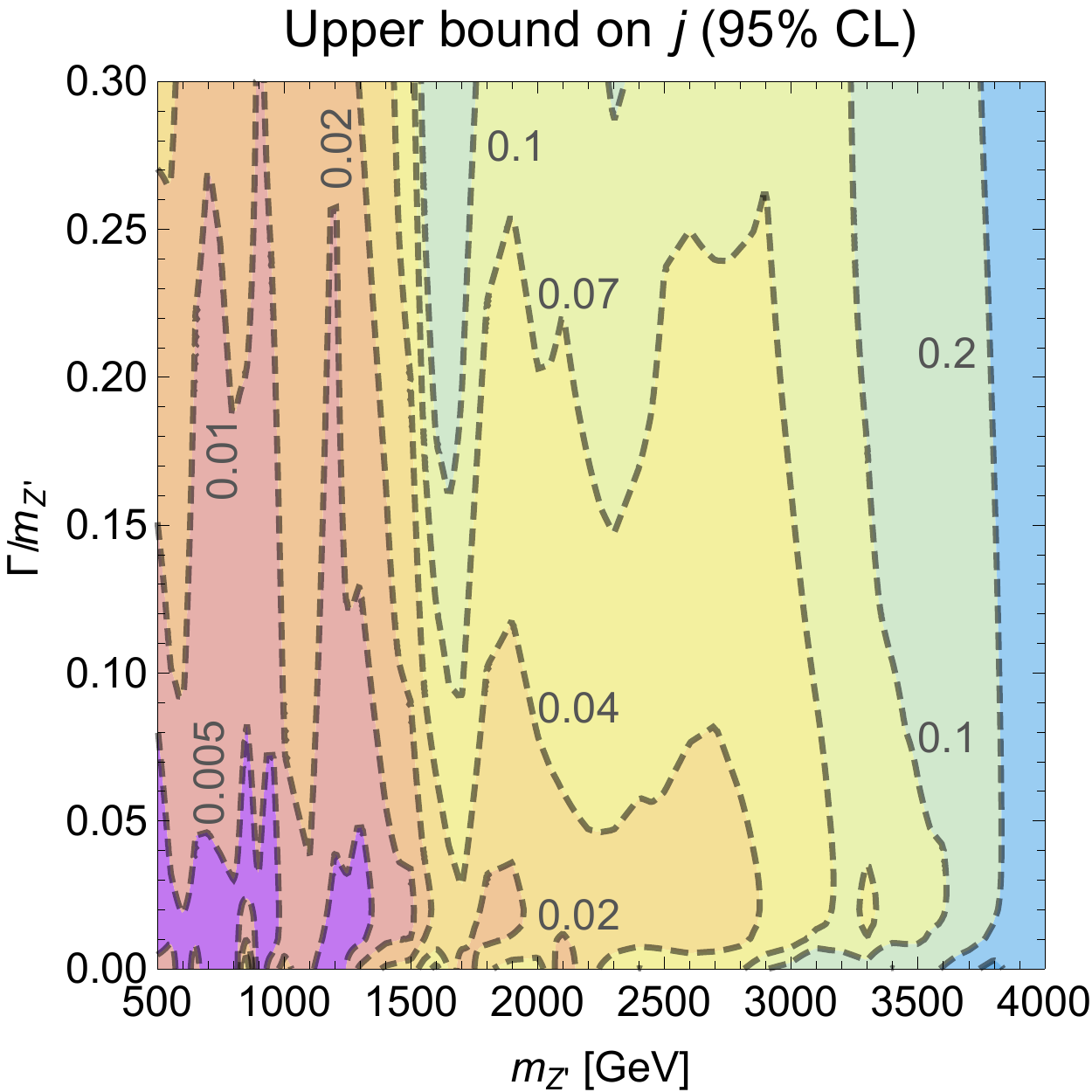}
  \caption{Bounds on $g_q$ (left) and $j \equiv g_q^2 \times \text{BR}(Z' \rightarrow jj)$ (right) from a combination of ATLAS and CMS dijet searches at 8 TeV and 13 TeV at 95\% confidence level as a function of the $Z'$ mass and width.}
  \label{fig:gq}
\end{figure}

While the resulting values of $g_{q,95\%}$ typically depend on $m_{Z'}$ in a non-monotonic way (due to different random fluctuations from bin to bin), the dependence on $\Gamma$ is typically very smooth, with larger values of $\Gamma$ corresponding to larger values of $g_{q,95\%}$ (i.e.\ weaker bounds). We can make use of this observation to interpolate between the values of $\Gamma$ considered in our simulation. Specifically, we find that it is possible to fit $g_{q,95\%}$ for fixed $m_{Z'}$ using a function of the form
\begin{equation}
g_{q,95\%}(m_{Z'},\Gamma_{Z'})^4 = a(m_{Z'}) \left(\frac{\Gamma_{Z'}}{m_{Z'}}\right)^{b(m_{Z'})} + c(m_{Z'}) \; ,
\label{poly}
\end{equation}
where the values of $a$, $b$ and $c$ are listed in table~\ref{tab:gq} in appendix~\ref{app:tab} as a function of $m_{Z'}$. We show $g_{q,95\%}$ as a function of $m_{Z'}$ and $\Gamma$ as obtained from the interpolation functions in the left panel of figure~\ref{fig:gq}.

For $m_{Z'} \lesssim 1.5\:\text{TeV}$ dijet constraints are able to exclude values of $g_q$ between 0.1 (for a narrow width) and 0.3 (for a broad width). For larger masses, these bounds become somewhat weaker and reach up to $g_{q,95\%} \approx 0.6$ for $m_{Z'} \sim 4\:\text{TeV}$ and $\Gamma / m_{Z'} > 0.2$. We observe that rather weak bounds are obtained for $m_{Z'} \approx 1.6\text{--}1.7 \:\text{TeV}$. The reason is that in this mass range all four experiments see an upward fluctuation in the data, so that the observed bound is weaker than the expected one (see also~\cite{Dev:2015pga,Das:2016akd}).\footnote{This pattern is driven by the ATLAS 8 TeV data set and is most pronounced for very broad widths. The largest preference for a non-zero contribution from a $Z'$ is found for $m_{Z'} = 1.7\:\text{TeV}$, $\Gamma / m_{Z'} = 0.3$ and $g_q = 0.55$. The local significance of this excess is 3.3$\sigma$ for ATLAS alone and 3.8$\sigma$ for the combination of all data sets.}

Since we have consistently treated $\Gamma$ and $g_q$ as independent parameters, our results can be applied to any $Z'$ model (with universal vector-like couplings to quarks) by applying the following procedure:
\begin{enumerate}
 \item For given $Z'$ mass and given couplings of the $Z'$ to all other particles in the theory, calculate the total decay width $\Gamma$.
 \item Look up $g_{q,95\%}$ for this value of $\Gamma$ and the assumed $Z'$ mass.
 \item If $g_{q,95\%}$ is larger than the assumed $Z'$-quark coupling, the parameter point is allowed. Otherwise, it is excluded at 95\% confidence level.
\end{enumerate}

While the procedure detailed above applies to $Z'$ models with universal vector couplings to all quarks, it is also possible for us to constrain more complicated models. For this purpose, we make use of the narrow-width approximation (NWA), which is valid as long as the width of the $Z'$ is small compared to its mass (typically $\Gamma / m_{Z'} < 0.3$). The NWA states that the cross section for the production of dijet events via a resonance factorises into the production cross section of the resonance and the probability for this resonance to decay into a pair of jets:
\begin{equation}
 \sigma(pp \rightarrow Z' \rightarrow jj) = \sigma(pp \rightarrow Z') \times \text{BR}(Z' \rightarrow jj) \; ,
\end{equation}
where $\text{BR}(Z' \rightarrow jj) = \Gamma(Z'\rightarrow jj)/\Gamma = 5 \, g_q^2 \, m_{Z'} / (4\pi \, \Gamma)$. In the model we consider the $Z'$ production cross section is proportional to $g_q^2$, with a constant of proportionality that depends on the $Z'$ mass and the centre-of-mass energy. Consequently, the dijet signal in each experiment is proportional to $g_q^2$ times the relevant branching ratio:
\begin{equation}
 \sigma(pp\rightarrow jj) \propto g_q^2 \times \text{BR}(Z' \rightarrow jj) \; .
\end{equation}
Indeed, this relation is also correct for the differential cross section, i.e.\ the shape of the dijet invariant mass distribution is independent of the coupling $g_q$ for fixed mediator mass and width.

This observation motivates a different way of presenting our results, namely to place an upper bound on the combination~\cite{Frandsen:2012rk}
\begin{equation}
j \equiv g_q^2 \times \text{BR}(Z' \rightarrow jj) \; .
\label{eq:j}
\end{equation}
As discussed above, $j$ is proportional to $g_q^4$ for fixed $\Gamma$. We can then calculate the upper bound on $j$ at 95\% confidence level, called $j_{95\%}$, by evaluating $j$ for $g_{j,95\%}$. We emphasise that it is perfectly acceptable for this calculation to yield a branching ratio larger than unity. In this case the conclusion would simply be that the experimental bounds cannot exclude any value of $g_q$ compatible with the chosen value of $\Gamma$.
 
Our results for $j_{95\%}$ are shown in the right panel of figure~\ref{fig:gq}. The advantage of this approach is that $j_{95\%}$ can also be used to constrain models beyond the one considered here. In particular, our analysis can be applied to the following cases:
\begin{itemize}
 \item For a $Z'$ with both vector ($g_q^V$) and axial ($g_q^A$) couplings to quarks, the production cross section for a $Z'$ is proportional to $(g_q^V)^2 + (g_q^A)^2$. In such a model, one should therefore calculate $\left[(g_q^V)^2 + (g_q^A)^2\right] \times \text{BR}(Z' \rightarrow jj)$ and compare the result to $j_{95\%}$ as shown in the right panel of figure~\ref{fig:gq}.\footnote{For a $Z'$ with purely axial couplings to quarks, one can also directly compare $g_q^A$ to $g_{q,95\%}$ shown in left panel of figure~\ref{fig:gq}.}
 \item The $Z'$ production is typically dominated by up and down quarks in the initial state. Consequently, for a $Z'$ with different couplings to the three generations, one can obtain an approximate bound by calculating $g_1^2 \times \text{BR}(Z' \rightarrow jj)$, where $g_1$ is the coupling to the first generation, and comparing the result to the bound on $g_q^2 \times \text{BR}(Z' \rightarrow jj)$ (i.e. $j_{95\%}$) shown in figure~\ref{fig:gq}.
\end{itemize}
For the convenience of the reader, we provide a plain text version of $j_{95\%}$ as a function of $m_{Z'}$ and $\Gamma$ in the supplementary material accompanying this paper. The advantage of these bounds is that they are independent of other interactions that may be present in a given $Z'$ model.

This discussion concludes the general analysis of dijet constraints. The remainder of the paper is dedicated to a specific application of the results shown above, which serves as an illustration for the procedure described above and uses this procedure in order to constrain a model of particular interest.

\section{Constraints on a leptophobic \texorpdfstring{$Z'$}{Z'} coupling to DM} \label{sec:DM}

We now show how the results from the previous section can be applied to a specific model. For this purpose, we consider a simple model of a leptophobic $Z'$ coupling to DM (see~\cite{Duerr:2013lka,Perez:2014qfa,Duerr:2014wra,Ohmer:2015lxa}), which is similar in spirit to the spin-1 s-channel simplified DM model discussed in refs.~\cite{Buchmueller:2014yoa,Harris:2014hga,Abdallah:2015ter,Abercrombie:2015wmb,Jacques:2015zha,DeSimone:2016fbz,Brennan:2016xjh}. Assuming DM to be a Majorana fermion $\psi$, the interactions between the SM and the dark sector are defined by the following Lagrangian:
\begin{align}
    \mathcal{L}_\text{kin} & =  \frac{i}{2} \bar{\psi} \gamma^\mu \partial_\mu \psi - \frac{1}{2} m_{DM} \bar{\psi} \psi -\frac{1}{4} F'_{\mu \nu} F'^{\mu \nu} + \frac{1}{2} m_{Z'}^2 Z'_{\mu} Z'^{\mu} \label{eq:lag1} \\
    \mathcal{L}_\text{int} & =- \frac{1}{2} g_{DM}^A Z'_\mu \bar{\psi} \gamma^\mu \gamma^5 \psi - g_q Z'_\mu \sum_q \bar{q} \gamma^\mu q \label{eq:lag2} \; .
\end{align}

The Majorana nature of the DM particle ensures that it can only have an axial coupling to the $Z'$, which significantly reduces constraints on the model from direct detection experiments. On the SM side, the couplings of the $Z'$ are assumed to be purely vectorial, which is consistent with the assumption that the $Z'$ couples neither to leptons nor to the SM Higgs~\cite{Kahlhoefer:2015bea}. We do not specify the additional dark Higgs necessary to generate the $Z'$ mass and the DM mass~\cite{Kahlhoefer:2015bea}, assuming that this particle is sufficiently heavy and sufficiently weakly mixed with the SM Higgs to be irrelevant for LHC phenomenology. While additional heavy fermions are needed to cancel anomalies, these can be colour-neutral and vector-like with respect to the SM gauge group, making them very difficult to observe at the LHC (see~\cite{Ekstedt:2016wyi} for a discussion of anomaly-free models).

The resulting decay widths are then given by\footnote{The pre-factor $1/(24 \pi)$ for the decay into DM results from the fact that there are two identical particles in the final state.}
\begin{align}
\Gamma(Z'\rightarrow q\bar{q}) & = \frac{m_{Z'} \, g_q^2}{4\pi} 
 \sqrt{1-\frac{4 m_q^2}{m_{Z'}^2}} \, \left(1 + 2 \frac{m_q^2}{m_{Z'}^2} \right) \label{eq:gamqq}  \\ \;
\Gamma(Z'\rightarrow \psi\psi) & = \frac{m_{Z'}}{24\pi} 
 (g^{A}_\text{DM})^2 \left(1-\frac{4 m_\text{DM}^2}{m_{Z'}^2}\right)^{3/2}  \label{eq:gamxx}\; ,
\end{align}
where we have assumed $m_{Z'} > 2 m_t, 2 m_\text{DM}$. The equations above enable us to calculate the total decay width $\Gamma$, which is required in order to apply the dijet bounds derived above.

\subsection{Bounds for fixed couplings}

The model introduced above has four free parameters (the two masses $m_{Z'}$ and $m_\text{DM}$ and the two couplings $g_q$ and $g_\text{DM}$). Since kinematic distributions at the LHC depend more sensitively on the masses than on the couplings, it is interesting as a first step to study LHC constraints for fixed couplings and varying masses. This approach is consistent with the most common way of presenting LHC constraints on DM simplified models. 

Note however that our model is not identical to any of the simplified models presently used by the ATLAS and CMS collaborations, because we consider a Majorana DM particle and a different coupling structure (in our model the $Z'$ has vector couplings to quarks and axial couplings to DM). Using a Majorana fermion instead of a Dirac fermion leads to an invisible width smaller by a factor of two giving slightly stronger bounds from dijet searches. The different coupling structure is not expected to significantly alter dijet bounds which typically depend on $(g_q^V)^2 + (g_q^A)^2$ (see above). It does, however, change the relic density constraint compared to the one obtained for the simplified models used by the experimental collaborations.

\begin{figure}[t]
  \centering
    \includegraphics[width=0.6\textwidth]{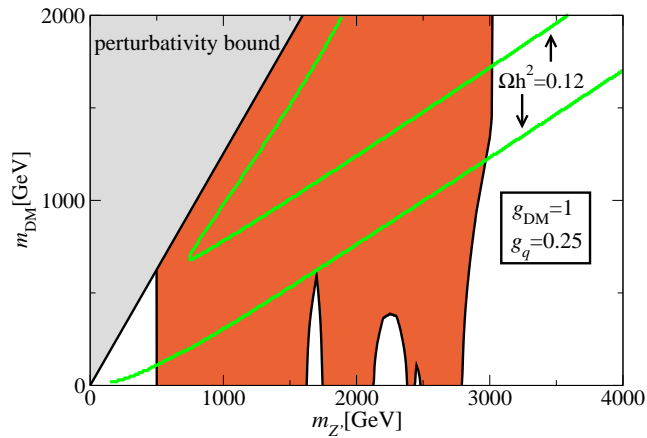}
  \caption{Excluded regions of parameter space in the mass-mass plane for fixed couplings, following the recommendations of the DM LHC working group~\cite{Boveia:2016mrp}. The region in red is excluded by our combined dijet analysis at the 95\% confidence level while the green lines represent parameter points which reproduce the observed relic density of DM in the Universe. In the grey region perturbative unitarity is violated. \label{fig:DMsimp}}
\end{figure}

Following the recommendations from~\cite{Abercrombie:2015wmb,Boveia:2016mrp}, we consider the case $q_q = 0.25$, $g_\text{DM} = 1$. For these couplings the width of the $Z'$ varies between 2.5\% (for $m_Z' < 2 m_\text{DM}, 2 m_t$) and 4.3\% (for $m_Z' \gg 2 m_\text{DM}, 2 m_t$) of its mass. For each combination of $m_{Z'}$ and $m_\text{DM}$, we calculate the width $\Gamma$ and then read off the largest allowed value for $g_q$ from figure~\ref{fig:gq}. Whenever $g_{q,95\%} < 0.25$, the parameter point is excluded by our combined dijet bounds. 
The results of this analysis are shown in figure~\ref{fig:DMsimp} with the dijet excluded regions shown in red. We find that $Z'$ masses between 500 GeV and 1600 GeV are excluded irrespective of the value of $m_\text{DM}$. For $m_{Z'}$ between 1600 GeV and 3 TeV, the model is excluded for heavy DM particles, such that the invisible branching ratio of the $Z'$ is small and decays into dijets dominate. These bounds are somewhat stronger than the ones found by the individual experiments due to our combined analysis.

For comparison, we also show the parameter region (in grey) where the coupling of the DM particle to the longitudinal component of the $Z'$ violates perturbative unitarity~\cite{Chala:2015ama,Kahlhoefer:2015bea} and  the masses for which the assumed couplings reproduce the observed DM relic abundance, $\Omega h^2 = 0.12$~\cite{Ade:2015xua}, calculated using \texttt{micrOMEGAs\_v4.1.8}~\cite{Belanger:2014vza}. Details on the relic density calculation can be found in the following subsection.

We find that there are only two regions where the relic density is compatible with dijet constraints: A low-mass region with $m_{Z'} < 500\:\text{GeV}$ and $m_\text{DM} < 150\:\text{GeV}$ and a high-mass region with $m_{Z'} > 3\:\text{TeV}$ and $m_\text{DM} > 1200\:\text{GeV}$.\footnote{We emphasise that in the low-mass region there may be additional dijet constraints from previous hadron colliders, as well as constraints from dijet resonances produced in association with SM gauge bosons~\cite{Chala:2015ama,Chiang:2015ika}. Moreover, this region of parameter space is tightly constrained by mono-X searches, in particular searches for jets in association with missing transverse energy~\cite{Khachatryan:2014rra,Aad:2015zva,Aaboud:2016tnv,CMS:2016tns}. These searches are very sensitive to $Z'$ masses below about 1 TeV, but lose sensitivity very quickly towards larger masses, where dijet constraints can still be sensitive.} This conclusion is, however, obviously dependent on our choice of couplings. For example, smaller values of $g_q$ would reduce the production cross section of the resonance, while larger values of $g_\text{DM}$ would increase the invisible branching ratio (provided $m_\text{DM} < m_{Z'} / 2$), so that dijet constraints could be significantly relaxed.

To study whether the intermediate mass range can be made compatible with dijet constraints for different choices of couplings, one could in principle repeat the analysis from above for many different combinations of $g_q$ and $g_\text{DM}$ (or simply scan the entire parameter space). Instead, we will take a more systematic approach and develop a new method that can be used to establish the compatibility of relic density constraints and dijet searches across the entire parameter space of our model (a similar comparison between LHC searches for dilepton resonances and the DM relic abundance in the context of gauged $B-L$ has been performed in~\cite{Okada:2016gsh}).

\subsection{Combining di-jet bounds and relic density}

Out of the four-dimensional parameter space of our model, we are particularly interested in those combinations of masses and couplings for which the thermal freeze-out of the DM particle can reproduce the relic abundance
\begin{equation}
\Omega h^2 = 0.1188\pm0.0010 \label{eq:relic}
\end{equation}
which is the result from Planck CMB observations combined with Baryon Acoustic Oscillations, supernova data and $H_0$ measurements~\cite{Ade:2015xua}.  We will approximate the relic density as $\Omega h^2=0.12$ in the rest of this work.

We emphasise that the relic density requirement can be relaxed if the dark sector consists of multiple components or if the thermal history of the Universe is non-standard. Nevertheless, it is certainly of interest to consider those parameters for which the simplest assumptions are already sufficient to match observations. If these parameters can be excluded experimentally, the model would require additional ingredients in the dark sector (such as additional annihilation channels, additional stable states or a mechanism to produce additional entropy after DM freeze-out), which by itself would be an important conclusion.

The remainder of this section focusses on how to reduce the parameter space of our model by imposing the relic density constraint. We first discuss some general aspects of the relic density calculation and then introduce a convenient set of free parameters that can be used to combine the relic density requirement with dijet constraints. Finally, we apply the dijet constraints from above to place bounds on the simple thermal freeze-out scenario.

To first approximation, we can obtain the relic density by calculating the cross section for DM annihilation into a pair of quarks, $\sigma_{\psi\psi \rightarrow q\bar{q}}$, and expanding the result in terms of the relative velocity $v$ of the two DM particles:
\begin{equation}
\sigma_{\psi\psi \rightarrow q\bar{q}} \, v \approx a + b \, v^2 + \mathcal{O}(v^4) \; .
\end{equation}
The relic abundance is then approximately given by
\begin{equation}
 \Omega \, h^2 \simeq 1.07 \times 10^9 \: \text{GeV}^{-1} \frac{x_\text{fo}}{M_\text{Pl} \sqrt{g_\ast}\left(a + 3b/x_\text{fo}\right)} \; ,
\end{equation}
where $x_\text{fo} \sim 20\text{--}30$ is the ratio of the DM mass and the freeze-out temperature and $g_\ast \sim 80\text{--}90$ is the number of relativistic degrees of freedom during freeze-out. For our model, we find $a = 0$ (due to the Majorana nature of DM) and
\begin{equation}
b = \frac{3\,(g_\text{DM})^2\, g_q^2}{12\pi} \frac{(m_q^2 + 2 \, m_\text{DM}^2)(1 - m_q^2 / m_\text{DM}^2)^{1/2}}{\left[(m_{Z'}^2 - 4 m_\text{DM}^2)^2 + (\Gamma \, m_{Z'})^2\right]} \; . \label{eq:b}
\end{equation}

For $m_\text{DM} \approx m_{Z'} / 2$, the denominator in eq.~(\ref{eq:b}) becomes very small and DM annihilation receives a resonant enhancement. In this case, an expansion in terms of the velocity of the two DM particles is insufficient for an accurate calculation of the relic density and numerical methods are needed. We therefore calculate the relic density using \texttt{micrOMEGAs\_4.1.8}~\cite{Belanger:2014vza}, including two modifications under the instruction of the authors (see appendix~\ref{app:micromegas}).\footnote{We thank Alexander Pukhov for providing us with these modifications and for his help in the implementation.} 

For given $m_{Z'}$, $m_\text{DM}$ and $g_\text{DM}$ we can then numerically determine the value of $g_q$ that is required to reproduce the observed relic density.\footnote{The width $\Gamma$ is determined internally by micrOMEGAs in a self-consistent way.} As long as $m_\text{DM}$ is well below the resonance region, i.e.\ $m_\text{DM} \ll m_{Z'}/2$, eq.~(\ref{eq:b}) implies that the annihilation cross section is proportional to $g_q^2 \, g_\text{DM}^2 \, m_\text{DM}^2 / m_{Z'}^4$. Therefore it is always possible to fix $g_q$ in such a way that the observed value of $\Omega \, h^2$ is matched and the solution is always unique. In the resonance region, the annihilation cross section is proportional to $g_q^2 \, g_\text{DM}^2 / \Gamma$, which is still a monotonic function of $g_q$ so that any solution is unique. However, since the expression $g_q^2 \, g_\text{DM}^2 / \Gamma$ remains finite for $g_q \rightarrow \infty$, it is possible that no solution exists. In short, as long as $g_\text{DM}$ is large enough, there will always be a unique value of $g_q$ that reproduces the relic abundance.

\begin{figure}[t]
\centering
  \includegraphics[width=0.47\linewidth]{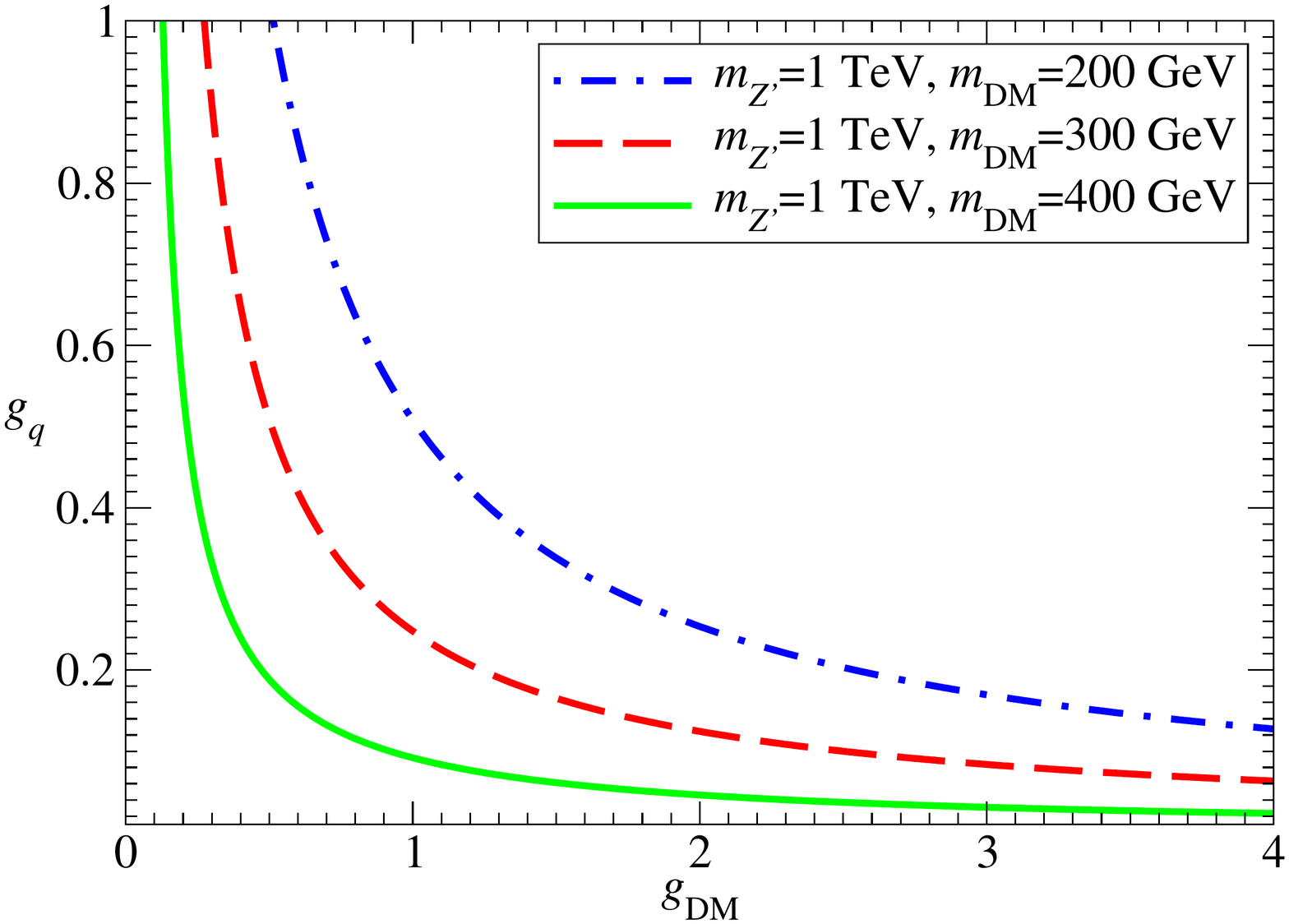}\qquad
  \includegraphics[width=0.47\linewidth]{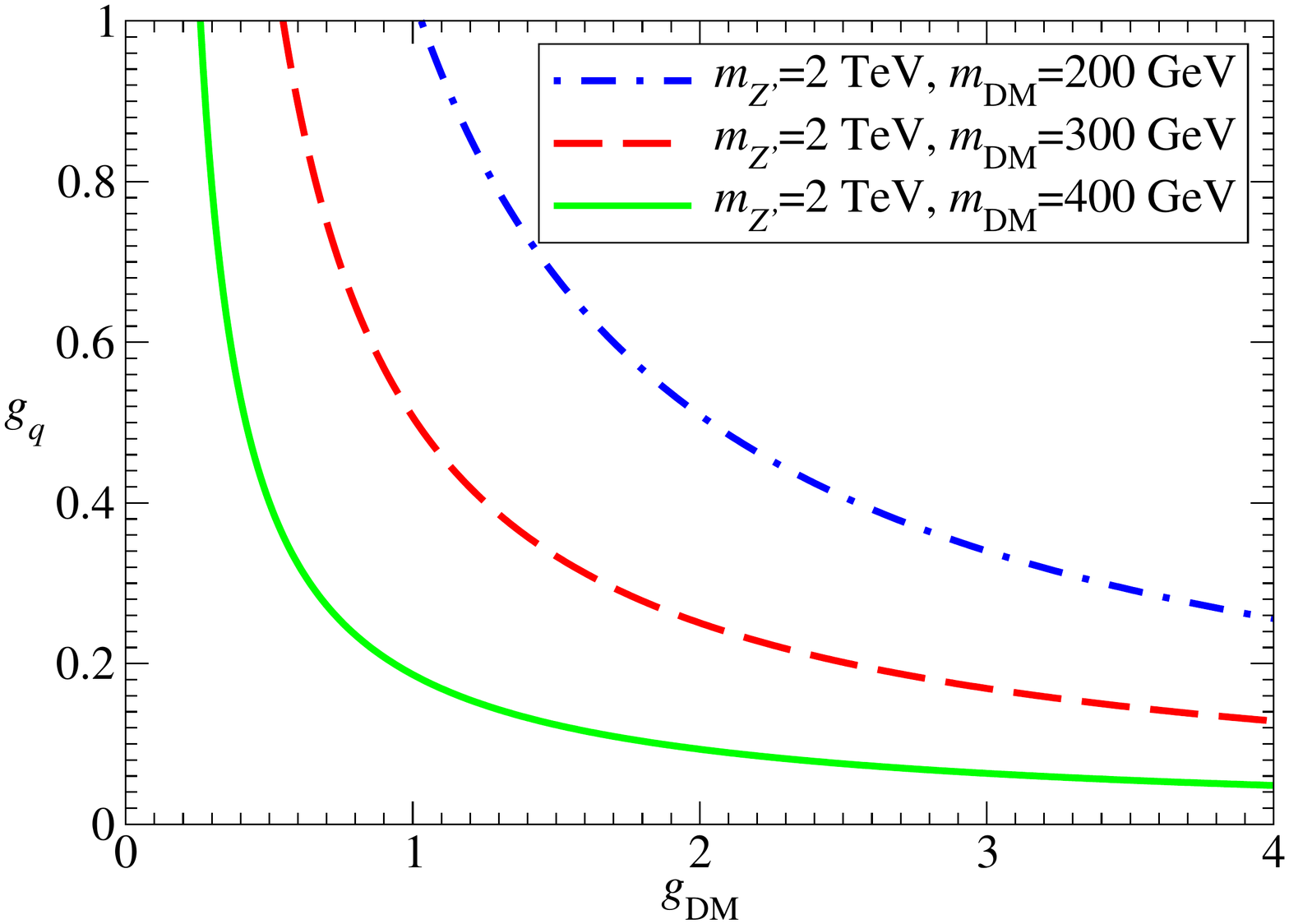}
\caption{Curves of constant relic density $\Omega h^2 = 0.12$ in the plane of the two couplings for fixed masses of the dark matter particle and a mediator mass of 1 TeV (left) and 2 TeV (right).}
\label{fig:relic}
\end{figure}

We therefore obtain a function $g_q(m_{Z'},m_\text{DM},g_\text{DM})$, which is illustrated in figure~\ref{fig:relic} as a function of $g_\text{DM}$ for various fixed values of $m_{Z'}$ and $m_\text{DM}$. The resulting curves have the following features:
\begin{enumerate}
 \item Since the annihilation cross section grows monotonically with the DM mass (for fixed couplings and mediator mass), the lines for different DM masses never cross, i.e.\ smaller values of $m_\text{DM}$ always require larger couplings.
 \item For sufficiently small DM masses, the curves are hyperbolas ($g_q \propto 1/g_\text{DM}$), whereas for larger values of $m_\text{DM}$, the curves are steeper at small $g_\text{DM}$ and flatter at large $g_\text{DM}$ due to the resonance effects discussed above.
\end{enumerate}
We will make use of these properties below to choose a particularly convenient set of free parameters for the analysis of our model.

\subsubsection{Relic density constraints for a fixed width}

Having constructed the function $g_q(m_{Z'},m_\text{DM},g_\text{DM})$ from the relic density requirement, one could now simply proceed to scan the remaining three-dimensional parameter space. One subtlety arises, however, from the fact that the width $\Gamma$~--- and therefore the bound from dijet constraints~--- depends on all three parameters in a non-trivial way. For example, for fixed $m_{Z'}$ and $m_\text{DM}$ one would naively expect stronger dijet constraints for smaller $g_\text{DM}$ corresponding to larger $g_q$ (implying both a larger production cross section of the resonance and a larger branching fraction into dijets). However, if at the same time $\Gamma$ increases, it is conceivable that dijet constraints are weakened sufficiently to evade experimental bounds and that in fact larger values of $g_q$ are less constrained than smaller couplings.

To avoid this complication, we take both $m_{Z'}$ and $\Gamma$ as free parameters. As shown in figure~\ref{fig:gq}, for fixed values of these two parameters we can always place an unambiguous upper bound on $g_q$. A second important advantage of this approach is that $m_{Z'}$ and $\Gamma$ are the two parameters that are most directly observable at the LHC. While the DM mass is very difficult to measure at the LHC and coupling constants can only be inferred in the context of a specific model, an observation of a new resonance in the dijet channel would immediately enable us to determine the mass and the width of the mediator from the invariant mass distribution.\footnote{This argument assumes that the width of the resonance is large compared to the detector resolution. Nevertheless, for a narrow resonance it is still possible to determine the mass and place an upper bound on the width.} To be able to directly interpret such an observation in the context of the present model, it therefore makes sense to construct all bounds in terms of these two most apparent observables.

\begin{figure}[t]
\centering
\includegraphics[width=0.47\linewidth]{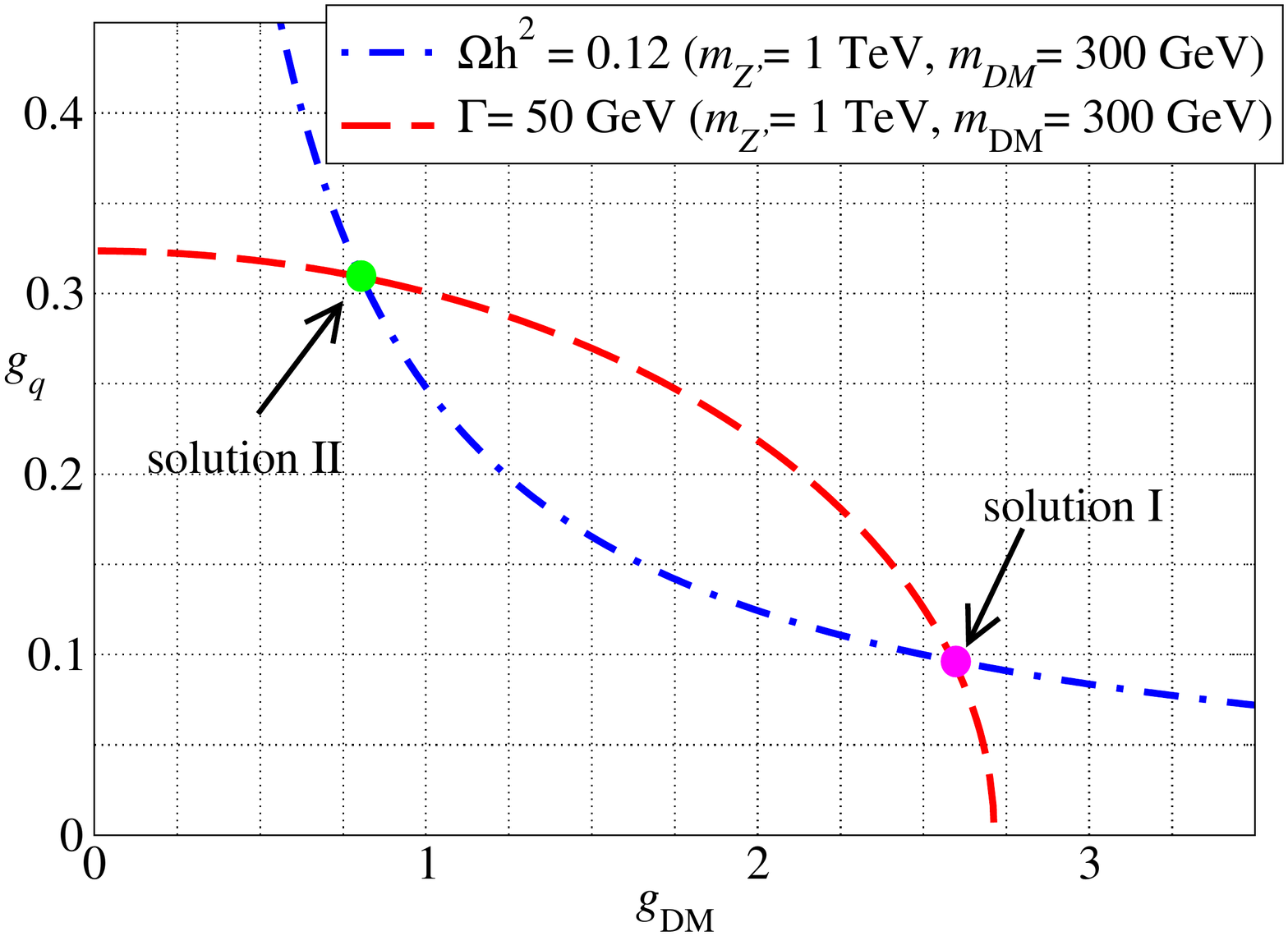}
\qquad
\includegraphics[width=0.47\linewidth]{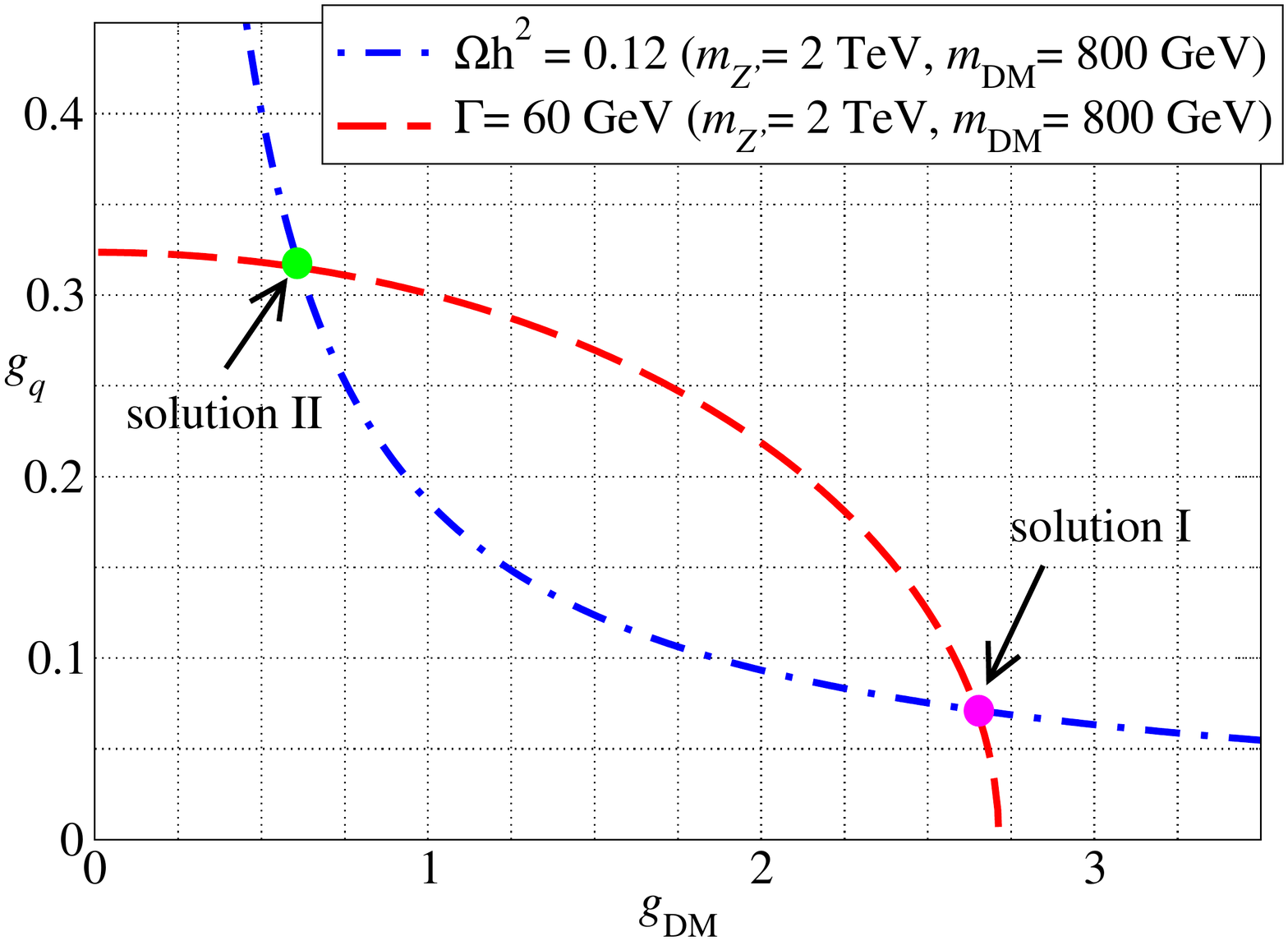}
\caption{Examples of how we find pairs of couplings that satisfy the relic density constraint (blue) for a given fixed width (red).}
\label{fig:cartoons}
\end{figure}

In order to treat the width $\Gamma$ as a free parameter, we need to determine those combinations of $m_\text{DM}$, $g_\text{DM}$ and $g_q$ that reproduce the observed relic density while at the same time matching the required width. For this purpose we first of all observe from eqs.~(\ref{eq:gamqq})--(\ref{eq:gamxx}) that for fixed $m_{Z'}$ and $m_\text{DM} < m_{Z'}/2$ the total width $\Gamma$ is an ellipse in the couplings.\footnote{For $m_\text{DM} \geq m_{Z'}/2$, the total width is independent of $g_\text{DM}$ and hence a straight line $g_q = \text{const}$.} We can now consider ellipses of constant width $\Gamma$ in the same $g_q$-$g_\text{DM}$-plane used in figure~\ref{fig:relic} to study the relic density constraints. Since the relic density curve is convex while the constant-width curve is concave, the two curves will have either exactly two intersects or zero intersects (neglecting those special cases where the two curves just touch at exactly one point). In other words, for fixed values of $m_{Z'}$, $\Gamma$ and $m_\text{DM}$, there is either no combination of $g_q$ and $g_\text{DM}$ that reproduces the relic density constraint or there are two separate solutions corresponding to the desired value of $\Gamma$. Whenever there are two solutions, we define Solution I to be the one with larger $g_\text{DM}$ (and therefore smaller $g_q$) and Solution II to be the one with smaller $g_\text{DM}$ (larger $g_q$). Two examples are shown in figure~\ref{fig:cartoons}. 

\begin{figure}[t]
  \centering
  \includegraphics[width=0.4\linewidth]{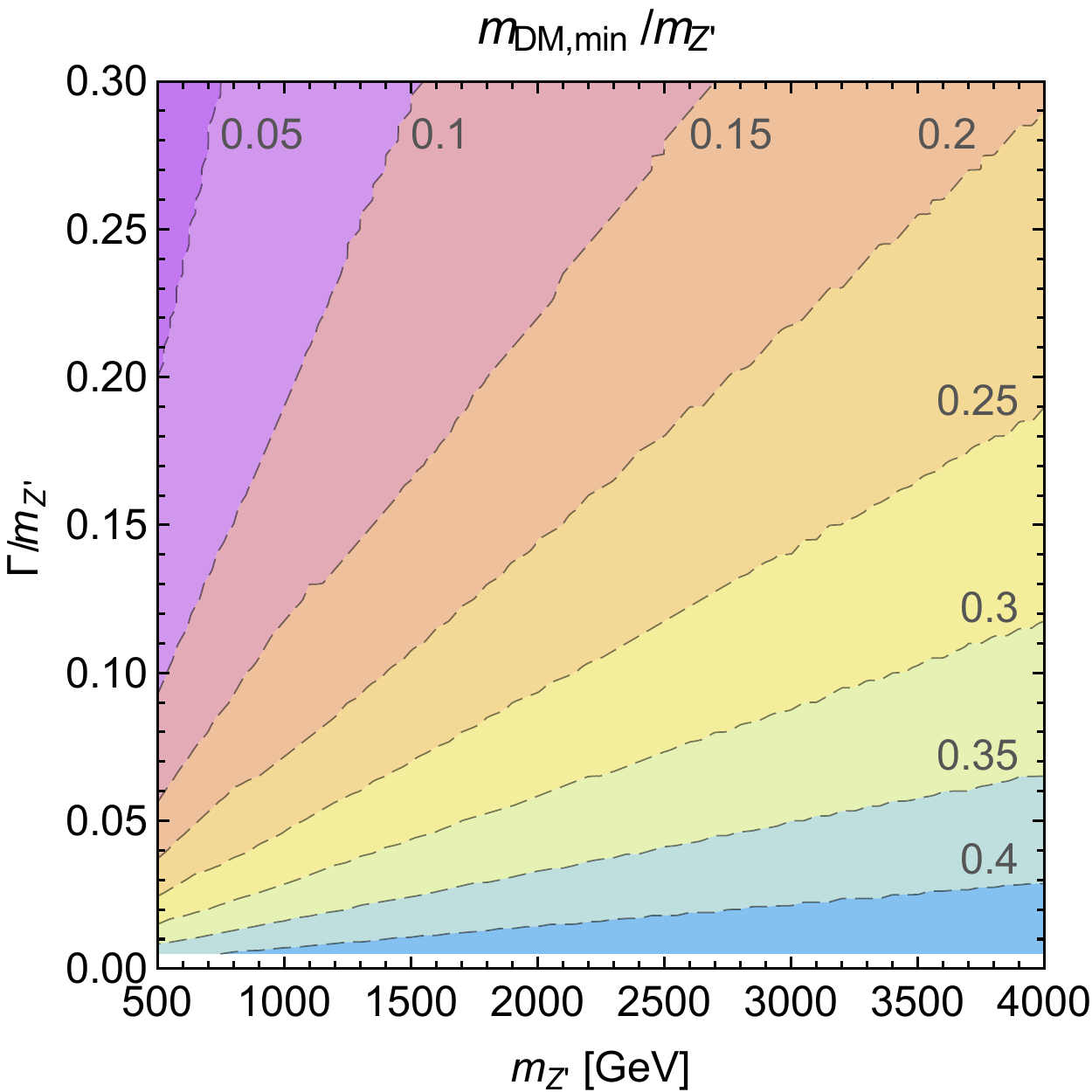}\qquad
  \includegraphics[width=0.4\linewidth]{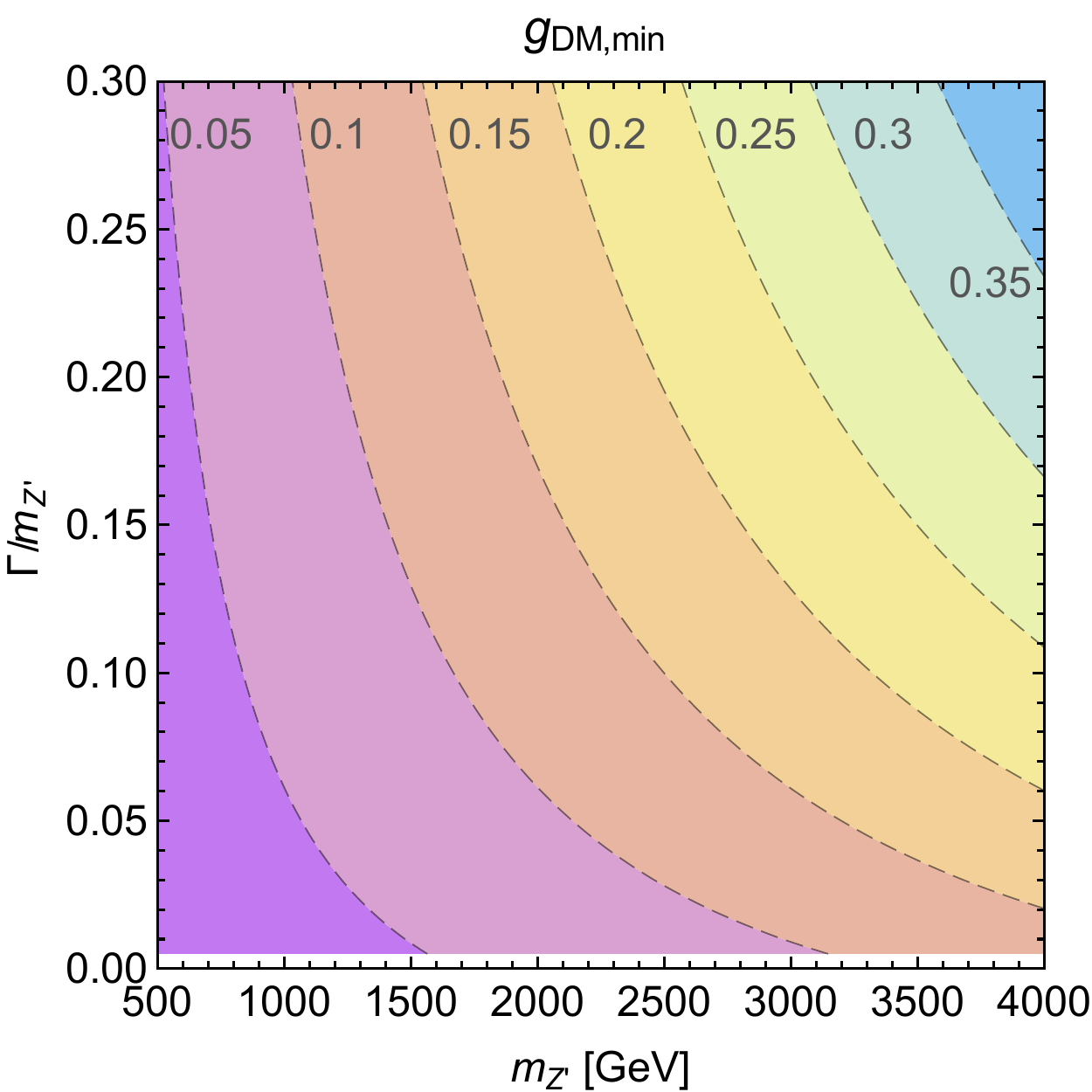}
  \caption{Left: The minimum value of the DM mass (in units of $m_{Z'}$) required in order to simultaneously satisfy the relic density constraint and reproduce the assumed $Z'$ width. For smaller DM masses, the relic density curve and the constant width curve do not intersect in the $g_\text{DM}$-$g_q$ plane (see figure~\ref{fig:cartoons}). Right: The smallest value of $g_\text{DM}$ that can reproduce the observed relic density and the assumed width. This value corresponds to Solution II for $m_\text{DM}$ slightly below $m_{Z'}/2$.}
  \label{fig:mDMmin}
\end{figure}

As noted above, increasing the value of $m_\text{DM}$ will shift the relic density curve towards smaller couplings. Conversely, the constant-width curve will be shifted to larger couplings (due to the larger phase-space suppression of $Z' \rightarrow \psi\psi$). This means that for each value of $m_{Z'}$ and $\Gamma$ there is a minimum value of $m_\text{DM}$, called $m_\text{DM,min}(m_{Z'}, \Gamma)$, such that there is no solution for $m_\text{DM} < m_\text{DM,min}$ and two solutions for $m_\text{DM} > m_\text{DM,min}$. Increasing $m_\text{DM}$ beyond $m_\text{DM,min}$ will shift Solution I to larger values of $g_\text{DM}$ and smaller values of $g_q$ and vice versa for Solution II. As $m_\text{DM}$ approaches $m_{Z'}/2$, Solution I will yield arbitrarily large values of $g_\text{DM}$ and thus ultimately violate the perturbativity bound $g_\text{DM} < \sqrt{4\pi}$. Solution II, on the other hand, will approach a small but non-zero minimum value of $g_\text{DM}$, called $g_\text{DM,min}$.\footnote{Note that in fact the resonant enhancement of the annihilation cross section is maximal (and hence the coupling $g_\text{DM}$ is minimal) for $m_\text{DM}$ slightly below $m_{Z'}/2$. We determine the DM mass corresponding to $g_\text{DM,min}$ numerically.} Figure~\ref{fig:mDMmin} shows both $m_\text{DM,min}$ and $g_\text{DM,min}$ as a function of $m_{Z'}$ and $\Gamma$.

With these considerations in mind, we can now eliminate either $m_\text{DM}$ or $g_\text{DM}$ in favour of $\Gamma$ and proceed with either $(m_{Z'},\,\Gamma,\, g_\text{DM})$, where $g_\text{DM,min} < g_\text{DM} < \sqrt{4\pi}$, or with $(m_{Z'},\, \Gamma,\, m_\text{DM})$, where $m_\text{DM,min} < m_\text{DM} < m_{Z'} / 2$. While in the first case we find a single value of $g_q$ for each set  $(m_{Z'},\,\Gamma\, g_\text{DM})$ compatible with the relic density constraint and consistent with the required width, the second case yields two separate solutions as discussed above. We discuss both possibilities below, as they each offer different physical insights.

\subsubsection{Dijet bounds on the DM coupling}

\begin{figure}[t]
\centering
\includegraphics[width=0.45\linewidth]{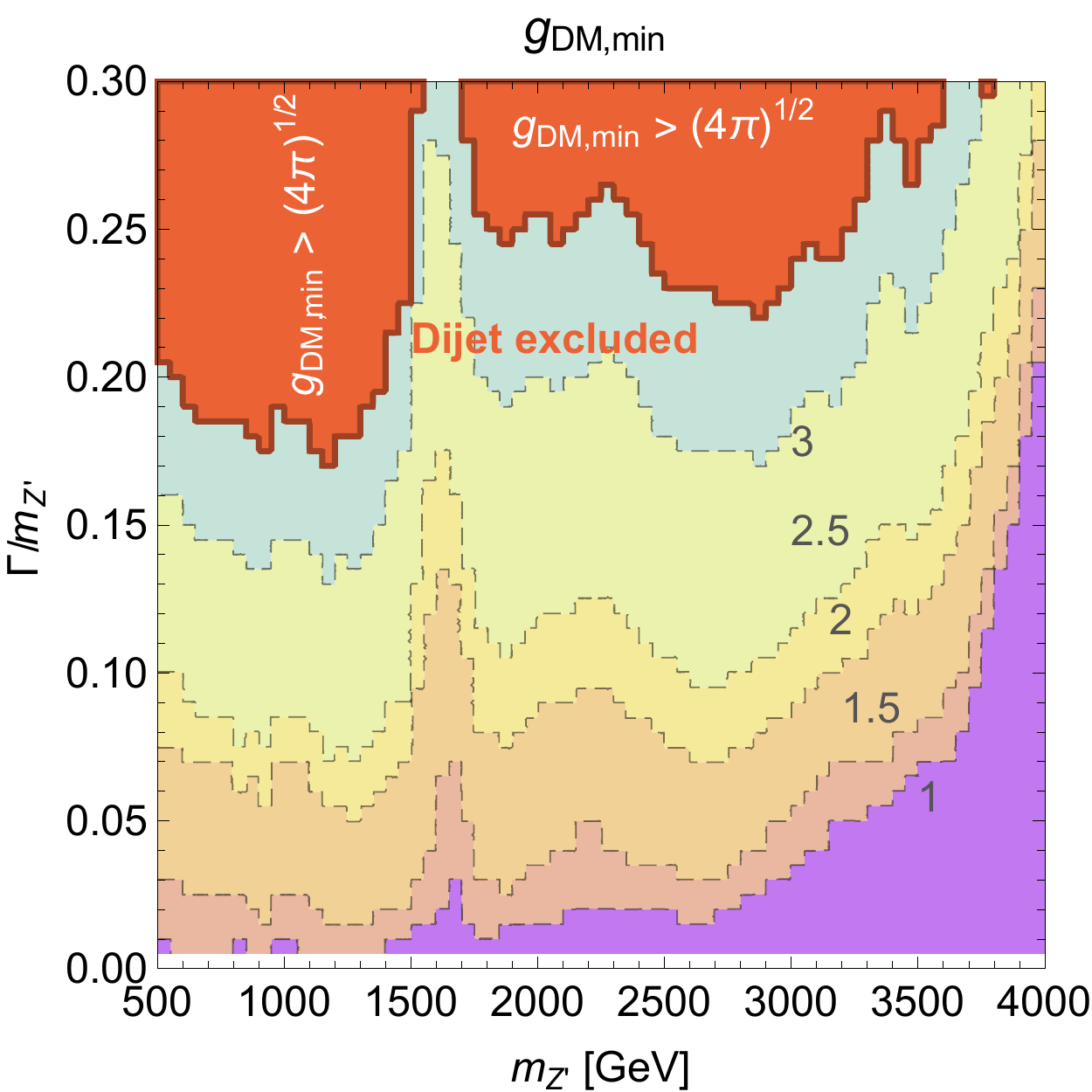}
\caption{Lower bound on the DM coupling $g_\text{DM}$ from the combination of the relic density constraint and LHC dijet searches. In the orange shaded region, $g_\text{DM,min}$ becomes non-perturbative, i.e.\ all perturbative values of $g_\text{DM}$ are excluded by LHC dijet searches.}
\label{fig:gDMmin}
\end{figure}

We have shown above that for fixed $m_{Z'}$ and $\Gamma$ smaller values of $g_\text{DM}$ correspond to larger values of $g_q$. Using figure~\ref{fig:gq} to read off the upper bound on $g_q$ from dijet searches thus allows us to place a lower bound on $g_\text{DM}$. This lower bound on $g_\text{DM}$ is shown in figure~\ref{fig:gDMmin}. Wherever no bounds from dijet searches can be placed, we simply show the smallest value of $g_\text{DM}$ for which the relic density curve and the constant-width curve intersect (called $g_\text{DM,min}$ above). If on the other hand the lower bound from dijet searches is so strong that it requires $g_\text{DM}$ to be larger than $\sqrt{4\pi}$, we conclude that it is impossible to find perturbative values of $g_q$ and $g_\text{DM}$ such that the width $\Gamma$ and the relic density can be reproduced without violating dijet constraints. The corresponding regions are shaded in orange in figure~\ref{fig:gDMmin}.

We observe that rather large values of $g_\text{DM}$ are required in order to avoid dijet constraints. While the consistency of the relic density requirement and the assumed width only required $g_\text{DM,min} \sim 0.1\text{--}0.3$ (see figure~\ref{fig:mDMmin}), dijet constraints require $g_\text{DM} > 1$ in almost the entire parameter space that we consider. For large $Z'$ width, even larger values of $g_\text{DM}$ are required in order to reduce the branching ratio of the $Z'$ into dijets. For $m_{Z'} \lesssim 1.5 \:\text{TeV}$ and $\Gamma/m_{Z'} \gtrsim 0.2$ as well as for $1.7\:\text{TeV} \lesssim m_{Z'} \lesssim 3.3\:\text{TeV}$ and $\Gamma/m_{Z'} \gtrsim 0.25$, all perturbative values of $g_\text{DM}$ that reproduce the relic abundance are excluded by dijet searches. For larger $Z'$ masses, LHC dijet searches lose sensitivity, but significant improvements in this mass range can be expected from upcoming runs of the LHC at 13 TeV.

\subsubsection{Dijet bounds on the DM mass}

Let us finally present our results from a complementary perspective by taking $m_\text{DM}$ as a free parameter and determining both $g_\text{DM}$ and $g_q$ from the relic density constraint and the requirement of a constant width. As discussed above, for each value of $m_\text{DM}$ we obtain two separate solutions, with Solution I (II) corresponding to larger (smaller) $g_\text{DM}$. For each of the two solutions, we can directly read off from figure~\ref{fig:gq} whether the parameter point is excluded by the combined dijet constraints that we have derived above. These exclusion limits in turn allow us to determine the allowed range of DM masses as a function of $m_{Z'}$ and $\Gamma$. We now discuss the two different solutions in turn.

As noted above, for Solution I (i.e.\ larger values of $g_\text{DM}$), the DM coupling increases with the DM mass. The requirement to have a perturbative coupling, $g_\text{DM} < \sqrt{4\pi}$, therefore gives an upper bound on $m_\text{DM}$, called $m_\text{DM,max}$. For some values of $m_{Z'}$ and $\Gamma$ we find that Solution I yields a non-perturbative value of $g_\text{DM}$ for all values of the DM mass, so for these combinations of $Z'$ mass and width only Solution II is of interest.

Conversely, for Solution I smaller DM masses correspond to larger values of $g_q$. Since large values of $g_q$ are excluded by LHC dijet searches, we can use the LHC bounds to place a lower limit on $m_\text{DM}$, called $m_\text{DM,min}$.\footnote{Note that even if LHC dijet searches are not constraining, there is always a lower limit on $m_\text{DM}$ from the requirement that the relic density curve and the constant-width curve intersect in the $g_q$-$g_\text{DM}$ plane.} The combination of the perturbativity requirement and LHC dijet searches therefore yield a range of permitted dark matter masses $[m_\text{DM,min},m_\text{DM,max}]$, which satisfy all of our constraints. In other words, for $m_\text{DM}$ in this range, it is possible to find values of $g_q$ and $g_\text{DM}$ that yield the observed relic abundance and are consistent with all other constraints that we consider.

\begin{figure}[t]
  \centering
  \includegraphics[width=0.45\linewidth]{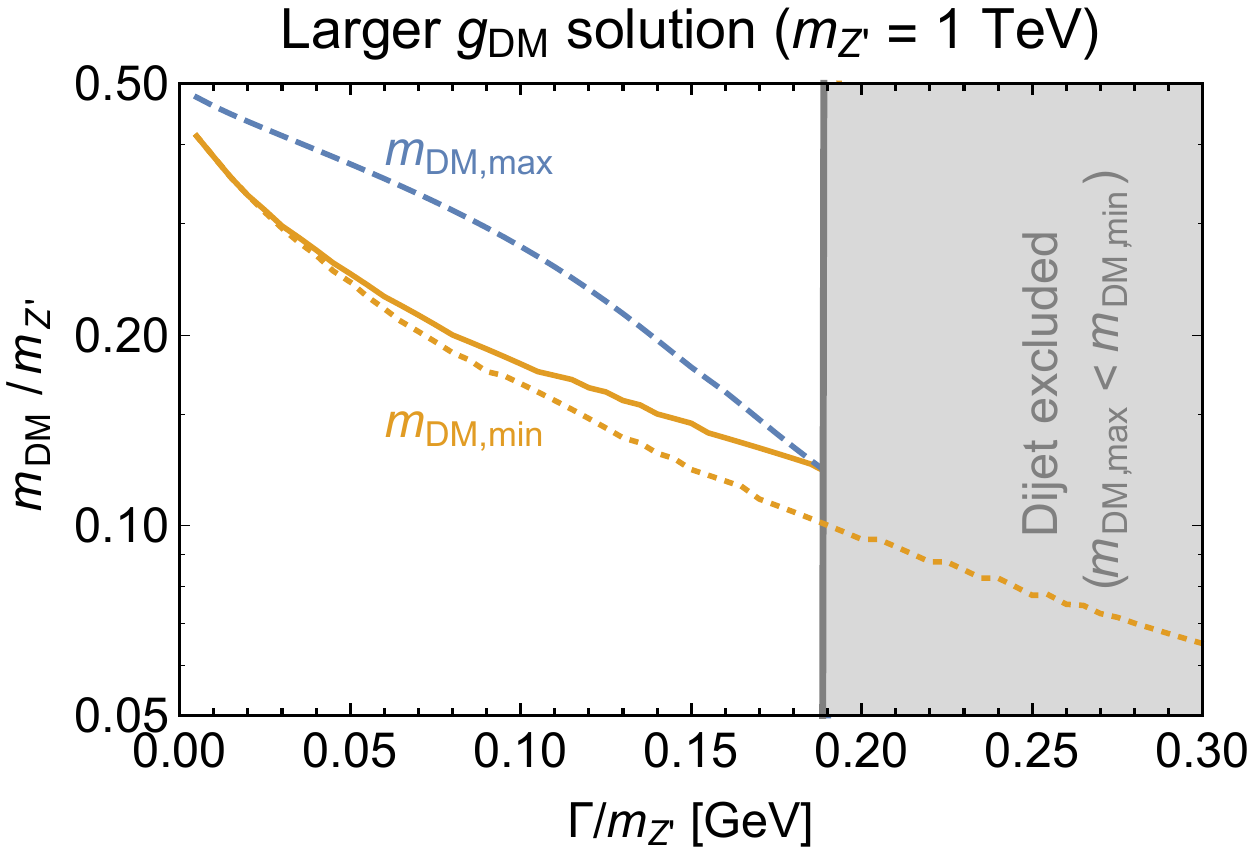}
  \caption{Maximum (blue, dashed) and minimum (orange) allowed value of the DM mass as a function of the mediator width for $m_{Z'} = 1\:\text{TeV}$ using Solution I (larger values of $g_\text{DM}$). The dotted orange line indicates the bound on $m_\text{DM,min}$ in the absence of LHC dijet constraints (see figure~\ref{fig:mDMmin}).}
  \label{fig:illustration}
\end{figure}

Figure~\ref{fig:illustration} shows one example, where we have fixed the $Z'$ mass to 1 TeV and show $m_\text{DM,min}$ (orange) and $m_\text{DM,max}$ (blue, dashed) as a function of $\Gamma$. To illustrate the impact of dijet searches, we also show the value of $m_\text{DM,min}$ that one obtains solely from the consistency of relic density and constant width (orange, dotted). For large values of $\Gamma$ we find that $m_\text{DM,max} < m_\text{DM,min}$, i.e.\ all perturbative solutions are excluded by LHC dijet searches. In the specific case under consideration, this occurs for $\Gamma / m_{Z'} \gtrsim 0.19$.

\begin{figure}[t]
  \centering
  \includegraphics[width=0.45\linewidth]{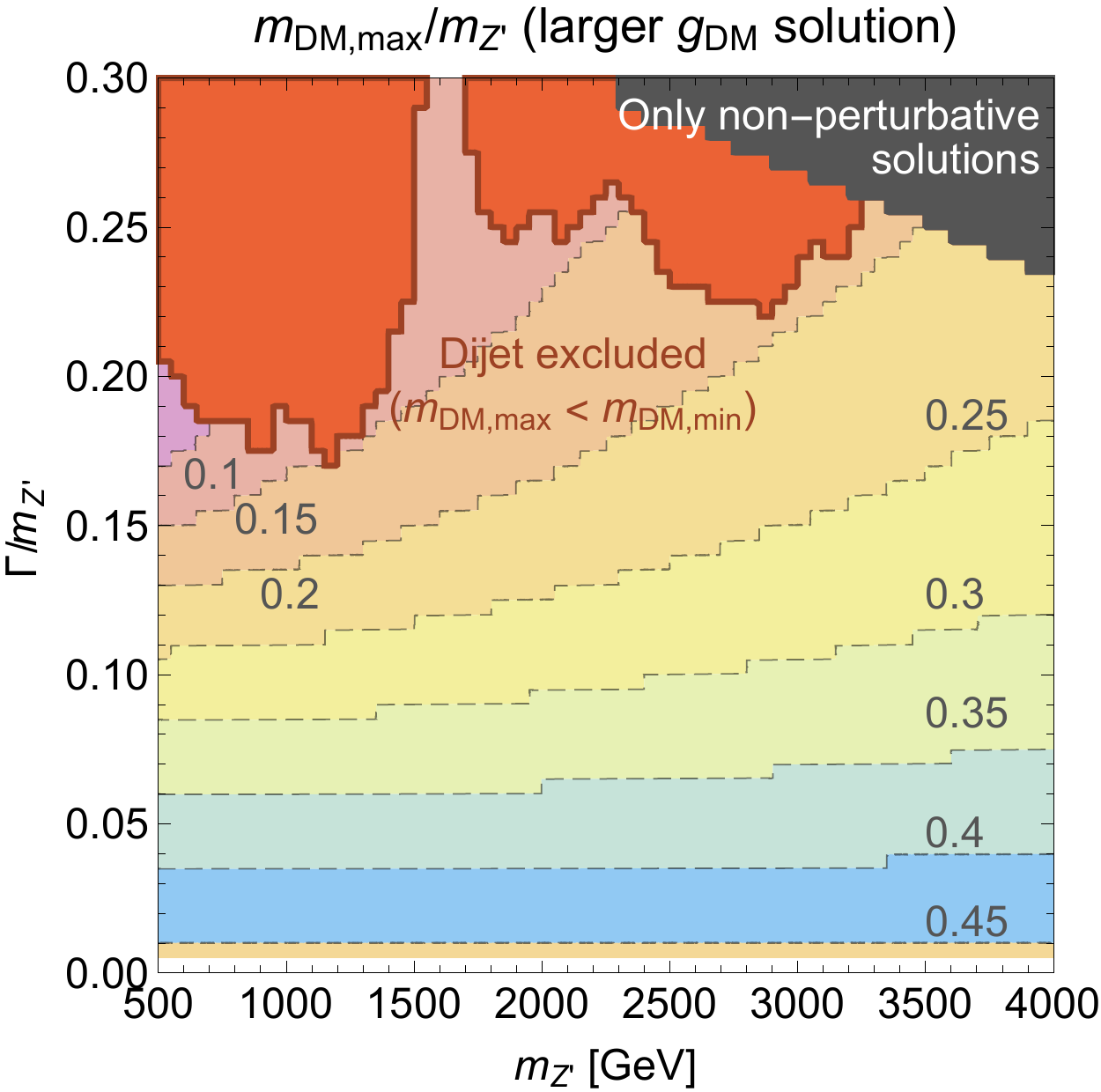}
\qquad
  \includegraphics[width=0.45\linewidth]{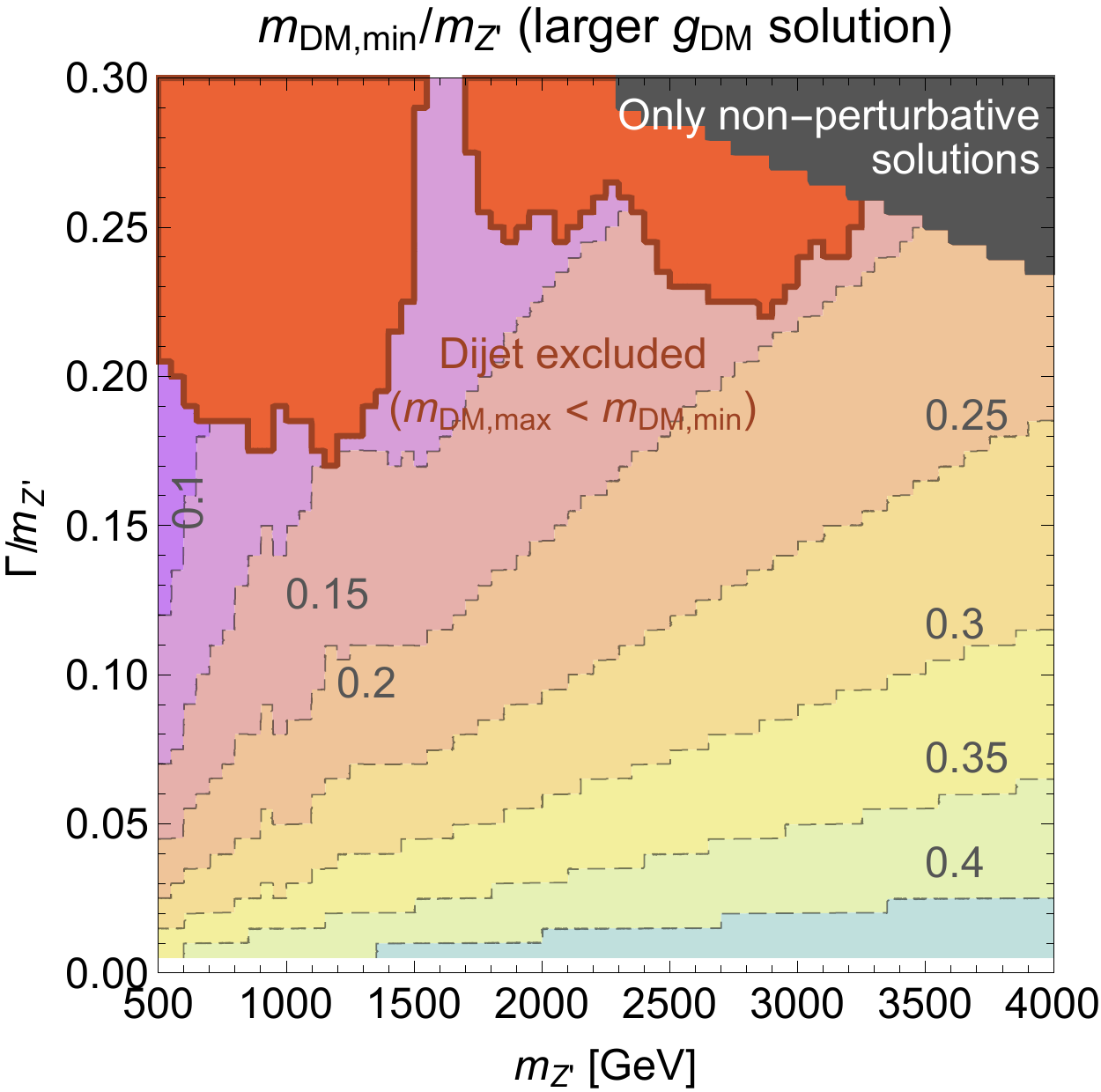}
  \caption{Maximum (left) and minimum (right) allowed value of the DM mass as a function of the mediator mass and width using Solution I (larger values of $g_\text{DM}$).}
  \label{fig:sol1}
\end{figure}

Figure~\ref{fig:sol1} shows the largest allowed DM mass (left) and the smallest allowed DM mass (right) as a function of $m_{Z'}$ and $\Gamma$. The plots can be read by picking a point on the plane (i.e.\ fixing $m_{Z'}$ and $\Gamma$) and then reading of $m_\text{DM,max}$ and $m_\text{DM,min}$ from the two panels to find the range of permitted dark matter masses $[m_\text{DM,min},m_\text{DM,max}]$ that satisfy all constraints. Those combinations of $m_{Z'}$ and $\Gamma$ for which $m_\text{DM,max} < m_\text{DM,min}$ are shaded in orange. The grey region indicates those combinations of $Z'$ mass and width for which no perturbative solutions can be found. As expected, the orange shaded region is identical to the one found in figure~\ref{fig:gDMmin}.

\begin{figure}[t]
  \centering
  \includegraphics[width=0.45\linewidth]{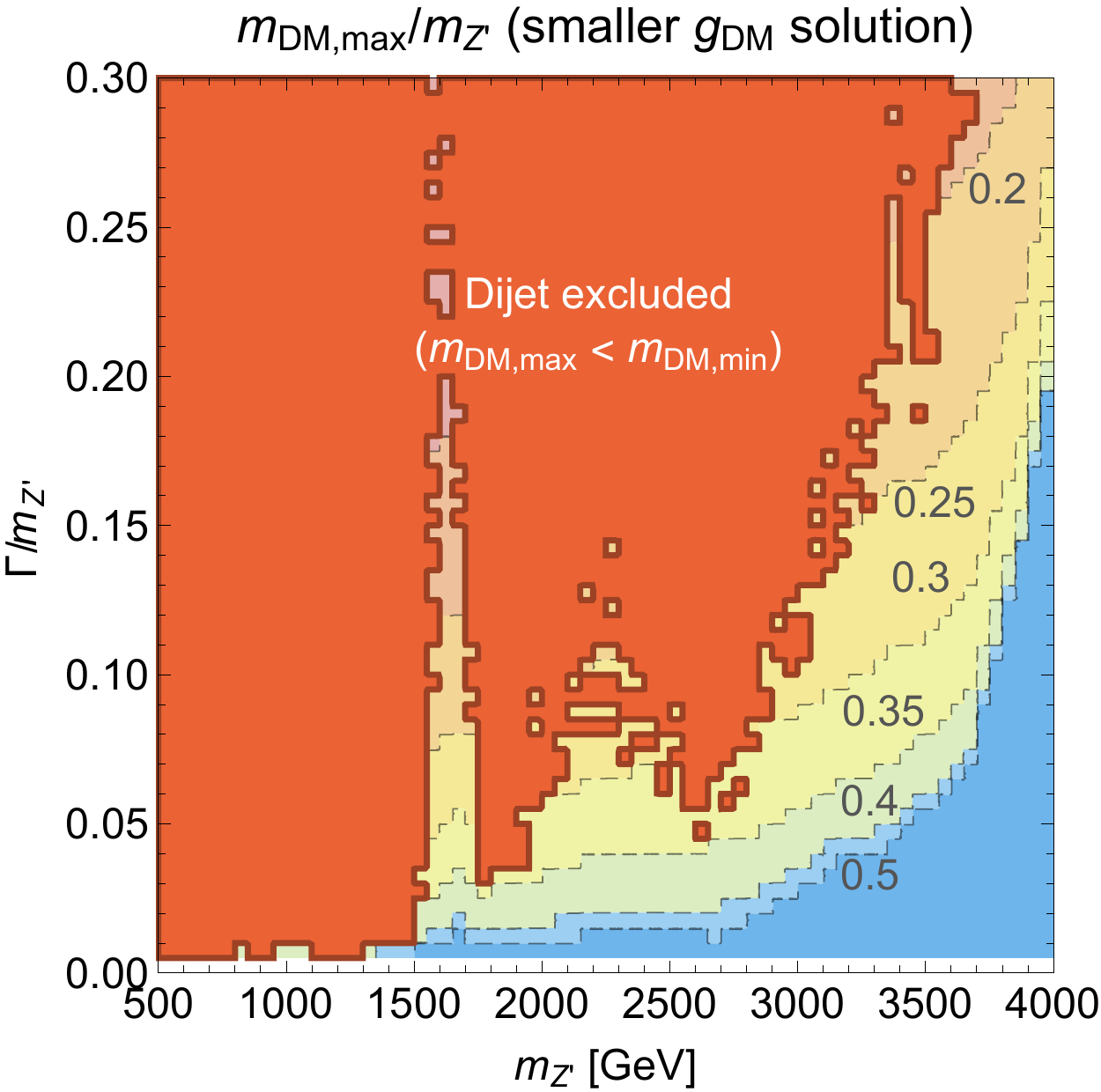}
\qquad
  \includegraphics[width=0.45\linewidth]{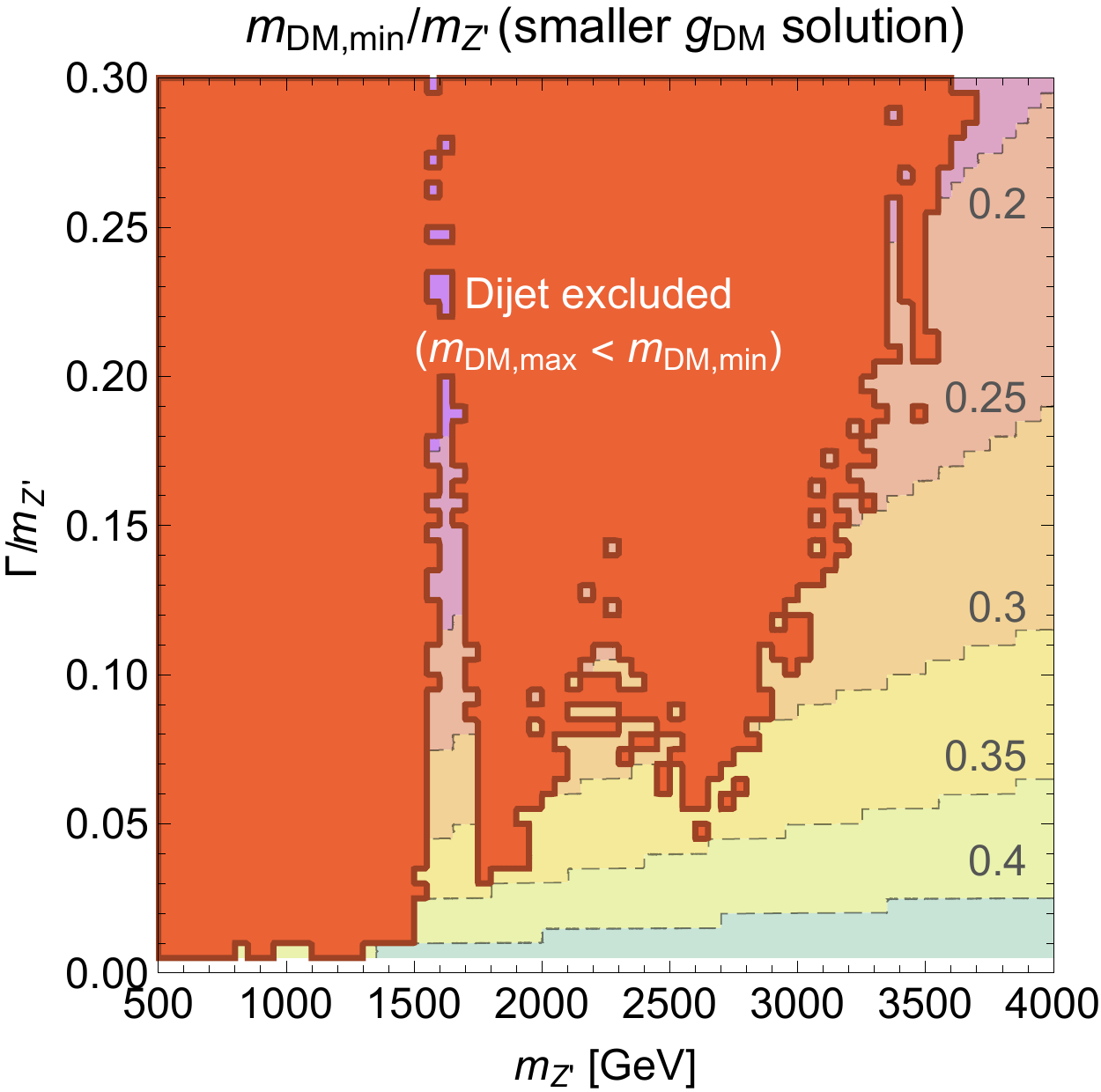}
  \caption{Maximum (left) and minimum (right) allowed value of the DM mass as a function of the mediator mass and width using Solution II (smaller values of $g_\text{DM}$).}
  \label{fig:sol2}
\end{figure}

Turning now to Solution II, we note that for this solution perturbativity constraints will typically be less important (because we consider smaller values of $g_\text{DM}$), while dijet constraints will be more important (because the corresponding values of $g_q$ are larger).\footnote{For the widths that we are considering, $\Gamma/M_{Z'} \leq 0.3$, $g_q$ is always less than unity, so we never run into problems with the perturbativity of $g_q$.} Compared to the previous solution, the situation is now reversed: The requirement of perturbativity may raise $m_\text{DM,min}$, while dijet constraints will lower $m_\text{DM,max}$. We show the maximum and minimum allowed DM masses for Solution II in figure~\ref{fig:sol2}. 

As expected, we find dijet constraints (shown in orange) to be significantly stronger than for Solution I. For large width, the entire range $500\:\text{GeV} \leq m_{Z'} \leq 3500\:\text{GeV}$ is excluded by dijet constraints. For $m_{Z'} \sim 1200\:\text{GeV}$ the dijet bounds extend down to very narrow resonances. As discussed above, dijet bounds are particularly weak around 1600 GeV, due to an intriguing upward fluctuation in the data. Finally, we note that we can always find a value of the DM mass such that Solution II corresponds to perturbative couplings (so the grey shaded region from figure~\ref{fig:sol1} is absent).

\section{Conclusions}

We have presented a combination of all available searches for dijet resonances at the LHC in the context of a generic $Z'$ model. Taking the width of the resonance and its coupling to quarks as independent parameters allows us to obtain constraints that apply irrespective of whether the $Z'$ decays exclusively into quarks or dominantly into other states. The results of this analysis, summarised in figure~\ref{fig:gq} and table~\ref{tab:gq}, are provided in such a way that they can be easily used to constrain a range of different models.

As a specific illustration of our approach, we have applied our constraints to a $Z'$ that couples to quarks and dark matter (DM), similar in spirit to a DM simplified model with a spin-1 $s$-channel mediator. It is straight-forward to map our constraints onto the parameter plane showing DM mass versus mediator mass for fixed couplings, which is conventionally used to present LHC results from missing energy searches. We show that for the typical choice of couplings ($g_\text{DM} = 1$, $g_q = 0.25$), dijet searches can exclude the range $500\:\text{GeV} < m_{Z'} < 3\:\text{TeV}$ for almost all values of the DM mass (see figure~\ref{fig:DMsimp}). These findings suggest that future searches for simplified DM models should focus on smaller values of $g_q$ and larger $g_\text{DM}$, which would relax constraints from searches for dijet resonances while still allowing for sizeable interactions between DM and quarks.

Finally, we have focussed on the special case that the $Z'$ mediates the interactions of DM and quarks responsible for thermal freeze-out, so that one of the parameters of the model can be eliminated by the requirement to reproduce the observed relic abundance. We have constructed a novel way of studying this set-up by making explicit the parameters that can be directly probed by searches for dijet resonances, i.e.\ the mass and the width of the $Z'$. The remaining free parameter can then be taken to be either the DM coupling (figure~\ref{fig:gDMmin}) or the DM mass (figures~\ref{fig:sol1} and \ref{fig:sol2}). We find that for very broad widths ($\Gamma/m_{Z'} \gtrsim 0.25$) and $Z'$ masses below about 3 TeV, LHC searches already exclude the possibility that the DM-quark interactions mediated by the $Z'$ are responsible for setting the DM relic abundance.

Furthermore, these figures provide a useful tool for interpreting future searches for dijet resonances at the LHC. Should an excess be seen in such a search, the mass and the width of the resonance can be determined from the data in a model-independent way. One can use these figures to look up whether the new state could conceivably act as the mediator into the dark sector. If a solution to the relic density requirement exists, the plots then provide the allowed ranges of the DM mass and coupling. Presently there is still ample room for such an interpretation, so there is much to be learned from the upcoming LHC data at 13 TeV.

\acknowledgments
 
We thank Antonio Boveia, Caterina Doglioni and Sungwon Lee for answering our questions on the various LHC dijet searches and Alexander Pukhov for significant help with micrOMEGAs. FK is supported by the German Science Foundation (DFG) under the Collaborative Research Center (SFB) 676 Particles, Strings and the Early Universe. MF and JH acknowledge support from the STFC.  MF and PT are funded by the European Research Council under the European Union's Horizon 2020 program (ERC Grant Agreement no.648680 DARKHORIZONS).

\appendix

\section{Tabulated bounds on \texorpdfstring{$g_q$}{gq}}
\label{app:tab}

We provide the numerical values of the dijet constraints obtained in section~\ref{sec:dijets} in table~\ref{tab:gq}.

\begin{table}[htb]
\renewcommand{\arraystretch}{0.83}
\caption{Numerical values of the fit to the constraint on the quark coupling as a function of width outlined in eq.~(\ref{poly}).\label{tab:gq}}
\begin{center}
  \begin{tabular}{|c|c|c|c|}
    \hline
    $M_{Z'}$[GeV]&$10\times a(M_{Z'})$ &$b(M_{Z'})$& $1000\times c(M_{Z'})$\\
    \hline
 500 &  2.4295 & 2.208 & 0.0588 \\
 550 &  1.3808 & 1.760 & 0.0712 \\
 600 &  0.7648 & 1.452 & 0.0000 \\
 650 &  0.5251 & 1.496 & 0.0584 \\
 700 &  0.4153 & 1.389 & 0.0000 \\
 750 &  0.4266 & 1.375 & 0.0000 \\
 800 &  0.4865 & 1.386 & 0.0000 \\
 850 &  1.2889 & 2.047 & 0.2376 \\
 900 &  0.3078 & 1.259 & 0.0000 \\
 950 &  0.7027 & 1.729 & 0.0540 \\
1000 &  0.5892 & 1.341 & 0.0743 \\
1100 &  0.4600 & 1.183 & 0.0000 \\
1200 &  0.3674 & 1.334 & 0.0000 \\
1300 &  1.4714 & 1.879 & 0.0809 \\
1400 &  1.8096 & 1.723 & 0.1545 \\
1500 &  4.4052 & 1.920 & 0.1901 \\
1600 & 13.7015 & 1.989 & 0.5733 \\
1700 &  5.4250 & 1.468 & 0.0000 \\
1800 &  5.1603 & 1.729 & 0.2606 \\
1900 &  4.8469 & 1.751 & 0.3736 \\
2000 &  4.7523 & 1.629 & 0.3762 \\
2100 &  3.3313 & 1.425 & 0.0000 \\
2200 &  3.9147 & 1.458 & 0.1891 \\
2300 &  4.9732 & 1.550 & 0.3758 \\
2400 &  4.9159 & 1.588 & 0.4994 \\
2500 &  3.3318 & 1.450 & 0.3996 \\
2600 &  3.5345 & 1.509 & 0.3922 \\
2700 &  3.9016 & 1.565 & 0.4383 \\
2800 &  3.2388 & 1.440 & 0.4402 \\
2900 &  2.6469 & 1.318 & 0.6291 \\
3000 &  3.0428 & 1.315 & 0.7375 \\
3100 &  3.5767 & 1.326 & 1.1560 \\
3200 &  2.6266 & 1.129 & 0.6442 \\
3300 &  4.0536 & 1.286 & 0.6851 \\
3400 &  6.1825 & 1.421 & 1.3730 \\
3500 &  3.7765 & 1.162 & 0.5939 \\
3600 &  5.0627 & 1.262 & 1.3046 \\
3700 &  6.5994 & 1.307 & 2.1257 \\
3800 &  6.8087 & 1.191 & 3.0885 \\
3900 &  5.6611 & 0.936 & 0.0000 \\
4000 &  9.5274 & 1.061 & 0.0000 \\
\hline
  \end{tabular}
  \end{center}
\end{table}

\section{Specific modifications of micrOMEGAs}
\label{app:micromegas}

In this appendix we detail two modifications to \texttt{CalcHEP}~\cite{Belyaev:2012qa}, which is used by \texttt{micrOMEGAs} to calculate the cross sections for DM pair annihilation. Both modifications are necessary in order to correctly treat the width of the $Z'$ close to the resonance (i.e.\ for $m_\text{DM} \approx m_{Z'}/2$).

The first modification is necessary to avoid numerical instabilities leading to kinks in the curves of constant relic density as a function of $g_\text{DM}$ and $g_q$ for fixed $m_\text{DM}$ and $m_{Z'}$ (see figure~\ref{fig:relic}). These kinks arise due to the way the Breit-Wigner (BW) propagator is implemented in CalcHEP. The standard BW distribution for a particle with mass $m$ and momentum $q$ is given by
\begin{equation}
|\mathcal{M}|^2 \propto \frac{1}{(q^2 - m^2)^2 + m^2 \Gamma^2} \label{eq:BW0} \; ,
\end{equation}
where the term involving $\Gamma$ removes the divergence as the particle becomes on-shell ($q^2 \rightarrow m^2$). However, since the width is a sum of diagrams of varying orders, its presence in eq.~(\ref{eq:BW0}) can spoil gauge invariance. For this reason CalcHEP implements the BW formula as a piecewise function over three regions: Formula \ref{eq:BW0} is used with a non-zero width for $|q^2 - m^2| < R\,m\,\Gamma$, where $R$ is an arbitrary number that is by default fixed to 2.7. For $|q^2 - m^2| > \sqrt{R^2 +1} \, m \, \Gamma$, on the other hand, the width in eq.~(\ref{eq:BW0}) is set to zero. In the intermediate region the width $\Gamma$ is replaced by a function of $q^2$ that interpolates between the two cases. For fairly large widths, as considered in the present work, this interpolation procedure can lead to kinks in the relic density calculation. To remove such kinks one can simply increase the value of $R$ from its default value by changing the value of the variable \texttt{BWrange}~\cite{calchep}. We have found that $R = 100$ is sufficient to remove the kinks in our plots.\footnote{Another effect that could lead to kinks in the relic density curves was due to a typographical mistake in \texttt{micrOMEGAs\_4.1.8}, which has been fixed in the most recent version.}

Furthermore, as pointed out in ref.~\cite{An:2012va}, the width $\Gamma$ of the resonance depends in general on the momentum transfer $q^2$, i.e.\ $\Gamma = \Gamma(q^2)$. For the case of narrow widths ($\Gamma/m \ll 1$), eq.~(\ref{eq:BW0}) gives a good approximation, because $\Gamma$ is only relevant for $q^2 \approx m^2$ and one can therefore approximate $\Gamma \approx \Gamma(q^2 = m^2)$. Since we consider widths as large as 30\% in this work, it is however appropriate to modify the BW formula in order to take the momentum dependence of the width into account. 

Following appendix A of~\cite{An:2012va}, we can approximately capture the momentum dependence by setting $\Gamma(q^2) = \frac{\sqrt{q^2}}{m} \Gamma(q^2 = m^2)$. This substitution yields
\begin{equation}
|\mathcal{M}|^2 \propto \frac{1}{(q^2 - m^2)^2 + \frac{(q^2)^2}{m^2} \Gamma(q^2 = m^2)^2}
\end{equation}
for the shape of the BW resonance. This modification can be implemented by editing the function \texttt{prepDen} used in the \texttt{CalcHEP} code.

\providecommand{\href}[2]{#2}\begingroup\raggedright\endgroup

\end{document}